\documentclass{article}

\usepackage{fullpage,amsfonts,amsmath,amsthm,graphicx}

\usepackage[usenames]{color}

\begin{document}

\definecolor{pink}{rgb}{1,0.08,0.45}
\newcommand{\refc}[1]{{\color{pink}{#1}}}

\def\a{\alpha}
\def\b{\beta}
\def\c{\chi}
\def\d{\delta}
\def\D{\Delta}
\def\e{\epsilon}
\def\f{\phi}
\def\F{\Phi}
\def\g{\gamma}
\def\G{\Gamma}
\def\k{\kappa}
\def\K{\Kappa}
\def\z{\zeta}
\def\th{\theta}
\def\Th{\Theta}
\def\l{\lambda}
\def\la{\lambda}
\def\m{\mu}
\def\n{\nu}
\def\p{\pi}
\def\P{\Pi}
\def\r{\rho}
\def\R{\Rho}
\def\s{\sigma}
\def\S{\Sigma}
\def\t{\tau}
\def\om{\omega}
\def\Om{\Omega}
\def\smallo{{\rm o}}
\def\bigo{{\rm O}}
\def\to{\rightarrow}
\def\E{{\bf Exp}}
\def\ex{{\mathbb E}}
\def\cd{{\cal D}}
\def\rme{{\rm e}}
\def\hf{{1\over2}}
\def\R{{\bf  R}}
\def\cala{{\cal A}}
\def\cale{{\cal E}}
\def\call{{\cal L}}
\def\cald{{\cal D}}
\def\calz{{\cal Z}}
\def\Fscr{{\cal F}}
\def\cc{{\cal C}}
\def\calc{{\cal C}}
\def\calh{{\cal H}}
\def\bk{\backslash}

\def\out{{\rm Out}}
\def\temp{{\rm Temp}}
\def\overused{{\rm Overused}}
\def\big{{\rm Big}}
\def\moderate{{\rm Moderate}}
\def\swappable{{\rm Swappable}}
\def\candidate{{\rm Candidate}}
\def\bad{{\rm Bad}}
\def\crit{{\rm Crit}}
\def\col{{\rm Col}}
\def\dist{{\rm dist}}
\def\poly{{\rm poly}}

\def\kwuumba{\ell^*}
\def\londara{\ell}

\newcommand{\Exp}{\mbox{\bf Exp}}
\newcommand{\var}{\mbox{\bf Var}}
\newcommand{\pr}{\mbox{\bf Pr}}

\newtheorem{lemma}{Lemma}
\newtheorem{theorem}{Theorem}
\newtheorem{corollary}[lemma]{Corollary}
\newtheorem{claim}[lemma]{Claim}
\newtheorem{remark}[lemma]{Remark}
\newtheorem{proposition}[lemma]{Proposition}
\newtheorem{definition}[lemma]{Definition}
\newtheorem{observation}[lemma]{Observation}

\newcommand{\limninf}{\lim_{n \rightarrow \infty}}
\newcommand{\proofstart}{{\bf Proof\hspace{2em}}}
\newcommand{\tset}{\mbox{$\cal T$}}
\newcommand{\proofend}{\hspace*{\fill}\mbox{$\Box$}}
\newcommand{\bfm}[1]{\mbox{\boldmath $#1$}}
\newcommand{\reals}{\mbox{\bfm{R}}}
\newcommand{\expect}{\mbox{\bf Exp}}
\newcommand{\he}{\hat{\e}}
\newcommand{\card}[1]{\mbox{$|#1|$}}
\newcommand{\rup}[1]{\mbox{$\lceil{ #1}\rceil$}}
\newcommand{\rdn}[1]{\mbox{$\lfloor{ #1}\rfloor$}}
\newcommand{\ov}[1]{\mbox{$\overline{ #1}$}}
\newcommand{\inv}[1]{\mbox{$\frac{1}{#1} $}}

\newcommand{\whp}{w.h.p.}

\date{\empty}

\title{
The solution space geometry of random linear equations
}

\author{
Dimitris Achlioptas\\
University of Athens\thanks{Research supported by an ERC IDEAS Starting Grant, an NSF CAREER Award, and a Sloan Fellowship.} \\
\and
Michael Molloy\\
University of Toronto\thanks{Dept.\ of Computer Science,
University of Toronto.  Research supported by an NSERC Discovery Grant.}
}

\maketitle

\begin{abstract}

We consider random  systems of linear equations over GF(2) in which every equation binds  {$k$} variables.  We obtain a precise description of the clustering of solutions in such systems. In particular, we prove that with probability that tends to 1 as the number of variables, $n$, grows: for every pair of solutions $\sigma, \tau$, either there exists a sequence of solutions {starting at $\sigma$ and ending at $\tau$ such that successive solutions have {Hamming} distance $O(\log n)$}, or every sequence {of solutions starting at $\sigma$ and ending at $\tau$ contains a pair of successive solutions with distance $\Omega(n)$}.  Furthermore, we determine precisely which pairs of solutions are in each category. {Key to our results is establishing the following high probability property of cores of random hypergraphs which is of independent interest. {Every} vertex not in the $r$-core of a random $k$-uniform hypergraph can be removed by a sequence of $O(\log n)$ steps, where each step amounts to removing one vertex of degree strictly less than $r$ at the time of removal.}

\end{abstract}

\newpage

\section{Introduction}\label{sintro}

In Random Constraint Satisfaction Problems (CSPs) one has a set of $n$ variables all with the same domain $D$ and a set of independently chosen constraints, each of which binds a randomly selected subset of $k$ variables. In the most common setting, both $D$ and $k$ are $O(1)$, while $m=\Theta(n)$. Two canonical examples are random $k$-SAT and coloring sparse random graphs. A fundamental quantity in the study of random CSPs is the so-called constraint density, i.e., the ratio of constraints-to-variables $m/n$.

There has been much non-rigorous evidence from statistical physics that for many random CSPs, if the constraint density is higher than a specific value, then all but a vanishing proportion of the solutions can be partitioned into exponentially many sets (clusters) such that each set is: (i) well-separated, i.e., has linear Hamming distance from all others, and (ii) in some sense, well-connected. The solution clustering phenomenon has been a central feature of the statistical physics approach to random CSPs and is  central to important algorithmic developments in the area, such as Survey Propagation~\cite{sp}. 

The mathematical studying of clustering began in~\cite{PhysRevLett.94.197205,daud} where it was shown that in random $k$-CNF formulas, above a certain density there exist constants $0 < \alpha_{{k}} < \beta_{{k}} < 1/2$  such that \whp\ no pair of satisfying assignments has distance in the range $[\alpha_k n ,\beta_k n]$. Let us say that two solutions are adjacent if they have Hamming distance 1 and consider the connected components under this notion of adjacency. In~\cite{fede} it was shown that above a certain density, there exist exponentially many  connected components of solutions and, moreover, in \emph{every one of them} the majority of variables are frozen, i.e., take the same value in all assignments in the connected component. 

Defining a cluster-region to be the union of one or more connected components, \cite{aminfede} proved that above a certain density, not only do exponentially many connected components exist, but there are exponentially many {cluster-regions} separated from one another by linear Hamming distance. Moreover, asymptotic bounds were given on the volume, diameter, and separation of these cluster regions. Later, in~\cite{aco}, it was shown for random $k$-SAT and random graph colouring {that when clustering occurs}, the emergent {cluster-regions} are also separated by large energetic barriers, i.e., that any path connecting solutions in different {cluster-regions} passes through value assignments violating linearly many constraints. {This picture of cluster-regions remains unchanged if one considers two solutions to be adjacent if they have Hamming distance $o(n)$. At the same time, though, it lends little information regarding the internal organization of cluster-regions, e.g., the connectivity of each such region.}

Until now, it has not been proven that \emph{any} random CSP model exhibits clustering into sets that are both well-separated and well-connected.  The main contribution of this paper is to prove that this phenomenon does indeed occur for random $k$-XOR-SAT, i.e., for systems of random linear equations over GF(2) where each equation contains precisely $k$ variables. We also obtain a precise description of the clusters. We remark that the cluster structure for $k$-XOR-SAT is much simpler than what is hypothesized for most CSPs, e.g., the clusters are all isomorphic and have the same set of {\em frozen variables} (see Section \ref{sec:intra}).
{Random} $k$-XOR-SAT has long been recognized as one of the most accessible of the fundamental random CSP models, in that researchers have managed to prove difficult results for it that appear to be far beyond our {current} reach for, e.g., random $k$-SAT and random graph coloring. {For example, the $k$-XOR-SAT satisfiability threshold was established by Dubois and Mandler\cite{dub} for $k=3$ and by  Dietzfelbinger {\em et al.}~\cite{cuc} for general $k$.} 

\subsection{Random systems of linear equations}

We consider systems of $m=O(n)$ linear equations over $n$ Boolean variables, where each equation binds a constant number of variables.
Clearly, deciding whether such a system has {satisfying assignments (solutions)} can be done in polynomial time by, say, Gaussian elimination. In fact, the set of solutions forms a subspace, so that the sum of two solutions is also a solution. At the same time, it seems that if one fails to exploit the underlying algebraic structure everything falls apart. For example, if the system is unsatisfiable, finding a value assignment $\sigma$ that satisfies as many equations as possible, i.e., MAX XOR-SAT, is NP-complete. Moreover, given a satisfiable system and an arbitrary $\sigma \in \{0,1\}^n$, finding a solution nearest to $\sigma$ is also NP-complete~\cite{berle}. Finally, random {systems of linear equations}
appear to be extremely difficult both for generic CSP solvers and for SAT solvers working on a SAT encoding of the instance. Indeed, very recent work strongly suggests that among a wide array of random CSPs, random $k$-XOR-SAT, {defined below}, is the \emph{most} difficult for random walk type algorithms such {as WalkSat}~\cite{young}.

{In random} $k$-XOR-SAT, which we study here, each equation binds exactly $k\ge 3$ variables (the case $k=2$ is trivial). To form the random system of equations $Ax =b$ we take $A$ to be the adjacency matrix of a random \mbox{$k$-uniform} hypergraph $H$ with $n$ variables and $m$ edges and $b \in \{0,1\}^m$ to be a uniformly random vector. It is straightforward to see, using e.g., Gaussian elimination, that if two systems have the same matrix $A$, {then their solution spaces are isomorphic
as $b$ ranges over vectors for which the solution space is not empty.} Since we will only be interested in properties of the set of solutions that are invariant under isomorphism, we will assume throughout that $b = \mathbf{0}$. As a result, throughout the paper we will be able to identify the system of linear equations with its underlying hypergraph. Regarding the choice of random $k$-uniform hypergraphs we will use both standard models $H_{{k}}(n,m)$ and $H_{{k}}(n,p)$, which respectively correspond to: including exactly $m$ out of the possible $\binom{n}{k}$ edges uniformly and independently, and including each possible edge independently with probability $p$. (Results transfer readily betwen the two models when $m= p \binom{n}{k}$.) Our corresponding models of random $k$-XOR-SAT are:
\begin{definition} $X_k(n,m),X_k(n,p)$ are the systems of linear equations over $n$ boolean variables
whose underlying hypergraphs are $H_k(n,m), H_k(n,p)$ and where we set  $b = \mathbf{0}$.
\end{definition}
The usual model for $k$-XOR-SAT differs from ours only in that it takes a uniformly random Boolean vector $b$.
As described above, these models are equivalent up to isomorphisms of the solution space, and hence
we can use our more convenient definition for the purposes of this paper. {We will say that a sequence of events $\mathcal{E}_n$ holds \emph{with high probability (\whp)} for such a system if $\lim_{n \to \infty} \Pr[\mathcal{E}_n] = 1$.}
We will analyze $X_k(n,p)$.  All our theorems translate to $X_k(n,m)$ where $m= p \binom{n}{k}$ using a standard argument.

We are interested in the range $p=\Theta(n^{1-k})$ which is equivalent to $m=\Theta(n)$.
{We note that} as $n \to \infty$, the degrees of the variables in such a random system tend to Poisson random variables with mean {$\Theta(1)$}. This implies that \whp\ there {will be} {$\Theta(n)$} variables of degree 0 and 1. Clearly, variables of degree 0 do not affect the satisfiability of the system. Similarly, if a variable $v$ appears in exactly one equation $e_i$, then we can always satisfy $e_i$ {by setting $v$ appropriately for any constant $b_i$.} Therefore, we can safely remove $e_i$ from consideration and only revisit it after we have found a solution to the remaining equations. Crucially, this removal of  $e_i$ can cause the degree of other variables to drop to 1. This leads us to the definition of the {core of a hypergraph.}

\begin{definition}
The {\em $r$-core} of a hypergraph $H$ is the {maximum} subgraph of $H$ in which every vertex has degree at least $r$.
\end{definition}

It is well known{~\cite{psw,mmcore,jhk}} that for every fixed $r \ge 2$, as $p$ is increased, $H_k(n,p)$ acquires a (massive) non-empty $r$-core suddenly, around a critical edge probability {$p=c^*_{k,r}/n^{k-1}$}.

Trivially, removing any vertex of degree less than $r$ {and all its incident edges} from $H$ does not change its $r$-core. Therefore, the $r$-core is the (potentially empty) outcome of the following procedure: repeatedly remove an arbitrary vertex of degree less than $r$ until no such vertices remain. {In the case of linear equations we will be particularly interested in  2-cores, as variables outside the 2-core can always be properly assigned.} 

\begin{definition}
The {\em 2-core system}   is the subsystem of linear equations induced by the 2-core {of the underlying hypergraph}, i.e., the set of equations whose variables all lie in the 2-core. A {\em 2-core solution} is a solution to the 2-core system. An {\em extension} of a 2-core solution, $\s$, is a solution of the entire system of linear equations that agrees with $\s$ on all 2-core variables.
\end{definition}

We will show that in the absence of a 2-core, {while the diameter of the set of solutions is linear, it is \whp\ possible to transform any solution to any other solution by changing $O(\log n)$ variables at a time. So, the set of solutions is not only well-connected but pairs of solutions exist at, essentially, every distance-scale. On the other hand, the emergence of the 2-core signals the onset of clustering, as now every pair of solutions is either very close with respect to the 2-core variables, or very far.}

\begin{theorem}\label{t1new}
For every {$k \ge 3$ and $c>c^*_{k,2}$}, there exists a constant $\a=\a(c,k)>0$ such that in $X_k(n,p={c/n^{k-1}})$, \whp\ every pair of solutions either disagree on at least $\a n$ 2-core variables, or on at most ${\xi}(n)$ 2-core variables, {for any} function ${\xi}(n)\rightarrow\infty$ arbitrarily slowly.
\end{theorem}

{We will} refine the picture of Theorem~\ref{t1new}, to prove that as soon as the \mbox{2-core} emerges, {unless two solutions agree on essentially all \mbox{2-core} variables, transforming one into another requires the simultaneous change of $\Omega(n)$ variables. To identify the relevant 2-core disagreements, we} need to define the following notion which is central to our work.
 \medskip

\begin{definition}\label{flippable_def}
A \emph{flippable cycle} in a hypergraph {$H$} is a set of vertices $S=\{v_1,\ldots,v_t\}$, where $t \ge 2$, 
{where the set of edges incident to $S$ can be ordered as $e_1, \ldots e_t$ such that}
each vertex $v_i$ lies in $e_i$ and in $e_{i+1}$ and in no other {edges of $H$}
 (addition mod $t$).
\end{definition}

\begin{remark}
Note that the vertices $v_1,\ldots,v_t$ must have degree exactly two in the hypergraph. The remaining vertices in edges $e_1,\ldots,e_t$ can have arbitrary degree and are \emph{not} part of the  flippable cycle.
\end{remark}

\begin{definition}\label{coreflippable_def}  A \emph{core flippable cycle} in  a hypergraph $H$ is a flippable cycle in the subhypergraph $H_0\subseteq H$ induced by the 2-core of $H$.
\end{definition}

Thus, in a core flippable cycle, the vertices $v_1,\ldots,v_t$ have degree exactly two \emph{in the 2-core}, but possibly higher degree in $H$. Note also that {$H$ may contain} flippable cycles outside the 2-core.  We will prove (Lemma~\ref{flip_cycles}(b)) that \whp\ the core flippable cycles are disjoint.

As discussed above, any 2-core solution can be readily extended to the remaining variables. Indeed, this can typically be done in numerous ways since the equations not in the 2-core are far less constrained, {e.g., a constant fraction of the equations outside the 2-core form hypertrees very loosely attached to the 2-core.} In order to understand the {emergence of the clustering of solutions, we will focus on whether we can} change the value of a 2-core variable without changing many other {2-core} variables.

If $\sigma$ is any 2-core solution then flipping the value of all variables in a core flippable cycle readily yields another solution of the 2-core, since every equation contains either zero or two of the flipped variables. It is not hard to show that  a random hypergraph often contains a handful of short core flippable cycles, implying that 2-core solutions may have Hamming distance $\Theta(1)$. At the same time, though, we will see (Lemma~\ref{flip_cycles}) that, for any ${\xi}(n)\rightarrow\infty$ arbitrarily slowly, the total number of vertices in core  flippable cycles \whp\ does not exceed ${\xi}(n)$, placing a corresponding upper bound on the distance between core solutions that differ only on flippable cycles.

In contrast, we will prove that \whp\ every pair of core solutions that differ on \emph{even one} {2-}core variable not in a flippable cycle,  differ in at least $\Omega(n)$ 2-core variables. In other words, flipping the {handful of variables in potential flippable cycles}, \whp\ is the {\em only} kind of movement between 2-core solutions that does not entail the simultaneous change of a massive number of variables.

The above indicates that the following is the appropriate definition of clusters in random $k$-XOR-SAT.

\begin{definition}
Two solutions are \emph{cycle-equivalent} if on the 2-core they differ only on variables in core flippable cycles ({while} they may differ arbitrarily on variables not in the 2-core).
\end{definition}

\begin{definition}\label{cluster_def}
The {\em solution clusters} of {$X_k(n,p={c/n^{k-1}})$} are the cycle-equivalence classes, i.e., two solutions are in the same cluster iff they are cycle-equivalent.
\end{definition}
Note that in the absence of a 2-core, this definition states that all solutions are in the same cluster. We can now state our main theorems in terms of connectivity properties of  clusters.

\begin{definition}
Two solutions $\sigma,\tau$ of a CSP  are \emph{$d$-connected} if there exists a sequence of solutions $\s,\s',\ldots,\t$ such that the Hamming distance of every two successive elements in the sequence is at most $d$.  A set $S$ of solutions is \emph{$d$-connected} if every pair $\s,\t\in S$ is $d$-connected.  Two solution sets $S,S'$ are
\emph{$d$-separated} if every pair $\s\in S, \t\in S'$ is \emph{not} $d$-connected.
\end{definition}

It appears that for many random CSP's, there is a constant $\a>0$ and a function $g(n)=o(n)$ such that if the constraint density is sufficiently large, then all but a vanishing proportion of the solutions can be partitioned into {\em clusters} $S_1,\ldots,S_t$ such that:
\begin{itemize}
\item Every $S_i$ is $g(n)$-connected. 
\item Every pair $S_i,S_j$ is $\a n$-separated. 
\end{itemize}

That is the sense in which we said earlier that each cluster is well-connected and that each pair of clusters is well-separated.

Our main theorems are that for $k$-XOR-SAT, the clusters we defined in Definition \ref{cluster_def} satisfy these conditions with $g(n)=O(\log n)$.  Note that for this particular CSP, the clusters contain \emph{all} the solutions, rather than  all but a vanishing proportion of them.

\begin{theorem}\label{t2}
For any constant ${c\neq c_{k,2}^*}$ and $k\geq 3$, there exists a constant {$\a=\a(c,k)>0$} such that in $X_k(n,p={c/n^{k-1}})$,  w.h.p.\ every pair of {clusters {is} $\a n$-separated.}
\end{theorem}

In stark contrast, we prove that clusters are internally very well connected.
\begin{theorem}\label{t1}
For any constant ${c\neq c_{k,2}^*}$  and $k\geq 3$, there exists a constant $Q=Q(c,k)>0$ such that in $X_k(n,p={c/n^{k-1}})$, w.h.p.\ {every cluster  {is} $Q\log n$-connected.}
\end{theorem}

Theorem~\ref{t1} is nearly tight due to the following.
\begin{observation}\label{olb}
W.h.p.\ every cluster contains a pair of solutions that are not $g(n)$-connected, for some $g(n)=\Omega(\log n/\log\log n)$.
\end{observation}
\begin{proof}
Consider  any solution $\s$ to the 2-core, and consider any two extensions $\s_0,\s_1$ of $\s$ to the entire {system} such that, for {some} 
non-core variable $v$, we have $\s_0(v)=0$ but $\s_1(v)=1$.  Then $\s_0,\s_1$ must differ in at least one additional variable in every equation containing $v$ implying that their Hamming distance is at least $\deg(v)+1$. 

{If $T$ is an acyclic {(tree)} component of the underlying hypergraph and $v$ is any vertex in $T$, then, clearly, $\s$ can be extended so that $v$ takes any desired value. Therefore, the maximum degree of any vertex in a tree component is a lower bound for $g(n)$. A tree component $T$ is a $d$-star if precisely one vertex in $T$ has degree $d$ and all other vertices have degree 1. Computing the second moment of the number of $d$-stars in a random hypergraph implies that \whp\ there exist $g(n)$-stars, where $g(n)=\Omega(\log n/\log\log n)$.}
\end{proof}

%


So, in a nutshell, we prove that before the 2-core emerges any solution can be transformed to any other solution {along} a sequence of successive solutions differing in $O(\log n)$ variables. In contrast, after the \mbox{2-core} emerges, the set of solutions shatters into clusters defined by complete agreement on the \mbox{2-core}, except for the handful of  variables in core flippable cycles: any two solutions that disagree on \emph{even one} 2-core variable not in a core flippable cycle, must disagree on $\Omega(n)$ variables. At the same time, solutions in the same cluster behave like solutions in the pre-core regime, i.e., one can travel arbitrarily inside each cluster by changing $O(\log n)$ variables at a time.

Our proof of Theorem~\ref{t1} is {algorithmic,} giving an efficient method to travel between any pair of solutions in the same cluster. {Indeed,} to prove Theorem \ref{t1},  we draw  heavily from the linear structure of the constraints to: (1) identify a set $B$ of \emph{free variables} such that the $2^{|B|}$ solutions in any cluster are determined by the $2^{|B|}$ assignments to $B$, (2) prove that we can change these free variables one-at-a-time, each time obtaining a new solution by changing only $O(\log n)$ other variables.

For $c<c_{k,2}^*$, there is no 2-core, and so all solutions belong to the same cluster.  For $c>c_{k,2}^*$,
but below the $k$-XOR-SAT satisfiability threshold,  the number of 2-core variables exceeds the number of 2-core equations by $\Theta(n)$ (see \cite{dub,cuc}), and the number of variables on core flippable cycles has expectation $O(1)$ (Lemma~\ref{flip_cycles}); it follows that w.h.p. there are an exponential number of clusters. So Theorems~\ref{t2},~\ref{t1} yield:
\begin{corollary}
For every $k\geq 3$ and $c$ below the $k$-XOR-SAT satisfiability threshold:
\begin{itemize}  
\item If $c<c_{k,2}^*$, then \whp\ the entire solution-set of $X_k(n,p=c/n^{k-1})$ is $O(\log n)$-connected.
\item If $c>c_{k,2}^*$, then \whp\ the  solution-set of $X_k(n,p=c/n^{k-1})$ consists of an exponential number of $\Theta(n)$-separated, $O(\log n)$-connected clusters.
\end{itemize}
\end{corollary}

{For $k \ge 3$,} the threshold for the appearance of a non-empty 2-core {was} determined in~\cite{mmcore,jhk} (see also \cite{cc}) to be:
\[c_{k,2}^*=
 \min_{\lambda > 0}
 \frac{(k-1)!\l}{\left(1-e^{-\lambda}\right)^{k-1}}
 \enspace .
 \]
For example, when $k=3$, an exponential number of clusters emerge  at $c=0.13...$ while the satisfiability threshold\cite{dub} is at $c=0.15...$. (These values correspond to $m/n=0.818...$ and $m/n=0.917...$ in the $X_k(n,m)$ model.) It is not clear what happens, in terms of clustering, at density $c=c_{k,2}^*$; see the remarks following Theorem~\ref{main_lemma}.

Our proof of Theorem~\ref{t2} easily extends to all \emph{uniquely extendable} CSPs.
\begin{definition}[\cite{cm}]
A constraint of arity $k$ is \emph{uniquely extendable} if for every set of $k-1$ variables and every value assignment to those variables there is precisely one value for the unassigned variable {that satisfies} the constraint.
\end{definition}
Linear equations over GF(2) and unique games are the two most common examples of uniquely extendable (UE) CSPs, but many others exist (see, eg. \cite{cm}). Clearly, any instance of a UE CSP $\Phi$ is satisfiable iff its 2-core is satisfiable. Thus, it is natural to define clusters analogously to XOR-SAT, i.e., two solutions are in the same cluster if and only if their 2-core restrictions differ only on core flippable cycles. Our proof of Theorem~\ref{t2} applies readily to any UE CSP, yielding a corresponding theorem, i.e., that {there exists $\a > 0$ such that if two solutions are not cycle-equivalent they are not $\a n$-connected} (see the remark following Proposition \ref{p2link}).  However, we do not know whether the analogue of Theorem \ref{t1} holds under this definition of clusters, i.e., whether it is possible to travel between cycle-equivalent solutions in small steps. Also, note that while in XOR-SAT changing all the variables in {\em any} flippable cycle results in another solution, this is not necessarily the case for every UE CSP $\Phi$.

Finally, we note that Theorem \ref{t1new} follows immediately from Theorem~\ref{t2} and the fact that w.h.p.\ there are fewer than ${\xi}(n)$ vertices on flippable cycles (Lemma~\ref{flip_cycles}).  Indeed, if two solutions differ on more than ${\xi}(n)$ 2-core variables, then they disagree on a variable that is not on a flippable cycle. Thus, they are in different clusters and so disagree on at least $\a n$ variables, by Theorem~\ref{t2}.  So the  paper focuses on proving Theorems~\ref{t2}~and~\ref{t1}.

{\bf Remark:} Ibrahimi, Kanoria, Kraning and Montanari~\cite{ikkm} have, independently,  obtained similar results to ours. Their definition of clusters is equivalent to ours, and they prove that the clusters are well-connected and well-separated. Their {cluster separation} result is equivalent to our Theorem \ref{t2}, but uses a different technique. Their internal connectivity result differs from our Theorem \ref{t1} in that (a) they prove that the clusters are $\mathrm{polylog}(n)$-connected rather than $O(\log n)$-connected, and (b) they {additionally} prove that the clusters exhibit a form of high conductance.  Again, their approach is  different from the one used here. To prove high conductance, they show that \whp\ the solution space contains a basis which is $\mathrm{polylog}(n)$-sparse, meaning that each vector in the basis has Hamming distance at most $\mathrm{polylog}(n)$ from $\mathbf{0}$. It is easy to see that the set of free variables $B$ that we choose in Section \ref{sec:intra} yields a $O(\log n)$-sparse basis. So our proof, along with Lemma 1.1 of \cite{ikkm} combine to yield a stronger conductance result by replacing ``$(\log n)^C$''  with ``$O(\log n)$'' in their Theorem 1.\medskip

\subsection{Cores of hypergraphs}
The main step in our proof of Theorem \ref{t1} is to prove a property of the non-2-core vertices in a random hypergraph. As this property is of independent interest, we prove it for non-$r$-core vertices for general $r\geq 2$.

For any integers $k\geq 2,r\geq 2$ {such that $r+k>4$,} the threshold for the appearance of a non-empty {$r$-core} in a $k$-uniform random hypergraph was determined in~\cite{mmcore,jhk} to be:
\begin{equation}\label{krthreshold}
c_{k,r}^*=\min_{\lambda > 0}
 \frac{(k-1)!\l}{\left[e^{-\lambda}\sum_{i = r-1}^{\infty} \l^i/i!\right]^{k-1}}
 \enspace .
 \end{equation}
For $r=k=2$, i.e., for cycles in graphs, the emergence of a 2-core is trivial as any constant-sized cycle has non-zero probability for all $c>0$. On the other hand, for $r+k>4$, any $r$-core has linear size \whp\ The threshold for the emergence of a 2-core of linear size in a random graph coincides with the threshold for the emergence of a giant component~\cite{pittel}, so we set $c^*_{2,2}=1$, consistent with the expression above after replacing $\min$ with $\inf$.

Recall that we can reach the $r$-core of a hypergraph by repeatedly removing any one vertex of degree less than $r$, until no such vertices remain. Consider {a} vertex $v$ not in the $r$-core, and consider the goal of repeatedly removing vertices of degree less than $r$ until $v$ is removed.  We prove that \whp\ for \emph{every} non-$r$-core variable $v$, this  can be achieved by removing only $O(\log n)$ vertices.

\begin{definition} An {\em $r$-stripping sequence} is a sequence of vertices that can be deleted from a hypergraph, one-at-a-time, {along with their incident hyperedges} such that at the time of deletion each vertex has degree less than $r$. A {\em terminal $r$-stripping sequence} is one that contains all vertices {outside the $r$-core;} i.e., a sequence whose deletion leaves {the $r$-core.}
\end{definition}

\begin{definition}
For any vertex $v$ not in the $r$-core, the {\em depth} of $v$ is the length of a shortest $r$-stripping sequence ending with $v$.
\end{definition}

\begin{theorem}\label{tcore}
For any integers $k\geq 2,r\geq 2$ and any constant {$c\neq c_{k,r}^*$}, let {$H = H_k(n,p=c/n^{k-1})$. There} exists a constant $Q = Q(c,k,r)>0$ such that \whp, every vertex $v$ in $H$ has depth at most $Q\log n$.
\end{theorem}

It is easy to show using standard facts about $r$-cores {of random hypergraphs} that for every constant $\e>0$, there is a constant $T=T(\e)$ such that w.h.p. all but $\e n$ of the non-core vertices have depth at most $T$. The challenge here  is to prove that \whp\ \emph{all} {non-core} vertices have depth $O(\log n)$.

\begin{remark} {\em The case $k=r=2$, i.e. the 2-core of a random graph, follows easily from previously known work.  The conclusion of Theorem~\ref{tcore} does not hold at $c=c^*_{2,2}=1$. (See the remarks following the statement of Theorem~\ref{main_lemma} below.) }
\end{remark}

\section{Related work}\label{sec:disc}

To get an upper bound on the random $k$-XOR-SAT satisfiability threshold, observe that the expected number of solutions in a random instance with $n$ variables and $m$ constraints is bounded by $2^n(1/2)^m \to 0$ if $m/n > 1$.
%
As one can imagine, this condition is not tight since variables of degree 0 and 1 only contribute fictitious degrees of freedom. Perhaps the next simplest necessary condition for satisfiability is $m_c/ n_c = \gamma_c \le 1$, where $n_c,m_c$ is the number of variables and equations in the 2-core, respectively. In~\cite{dub} Dubois and Mandler proved that, for $k=3$, this simple necessary condition for satisfiability is also sufficient by proving that for all $\gamma_c < 1$, the number of core solutions is strongly concentrated around its (exponential) expectation. Thus, they determined the satisfiability threshold for 3-XOR-SAT. Dietzfelbinger {\em et al.}~\cite{cuc} modify and extend the approach of \cite{dub} to determine the satisfiability threshold for general $k$.  A full version of \cite{dub} has not been published, but a  proof for all $k\geq 3$ appears in \cite{cuc}.

M\'{e}zard {\em et al.}~\cite{mez} were the first to study clustering in random $k$-XOR-SAT. Specifically, they defined the clusters by saying that two solutions are in the same cluster iff they agree on {\emph{all}} variables in the 2-core. They proved that there exists a constant $\gamma > 0$ such that for any $\theta \in(0,\gamma)$ {and any integer $z=\theta n+o(n)$, \whp\ no two solutions differ on exactly $z$} variables in the 2-core. {Based on this fact, they claimed that the clusters they defined are $\Omega(n)$-separated, i.e., that every pair of solutions in different clusters is not $\g n$-connected. As we have already seen, this is false since it does not account for the effect of core flippable cycles. Performing the analysis of solutions that differ on $o(n)$ variables is what allows us to establish that 2-core solutions which differ on $o(n)$ variables must differ {\em only} on core flippable cycles. Indeed, this is the most difficult part of our proof of Theorem~\ref{t2}.

M\'{e}zard {\em et al.}~\cite{mez}  also gave a heuristic argument that if $\sigma$ is any solution and $v$ is a non-core variable, then there exists a solution $\sigma'$ in which $v$ takes the opposite value from the one in $\sigma$ such that the distance between $\sigma$ and $\sigma'$ is $O(1)$. From this they concluded that clusters are well-connected.} Regarding internal connectivity, the clusters of~\cite{mez} are, indeed, well-connected, i.e., the analogue of Theorem~\ref{t1} holds for them, since they are subsets of the clusters defined in this paper. However their proof of this fact is flawed; it implies that their clusters are $O(1)$-connected, which is not true by the same argument used as for Observation \ref{olb}.  Proving that the $k$-XOR-SAT clusters are well-connected was later listed as an open problem in \cite{mmbook}.

Finally, as described above, Ibrahimi, Kanoria, Kraning and Montanari~\cite{ikkm} have, independently,  obtained similar results to ours.

\section{Proof Outline}

\subsection{Theorem~\ref{t2}: Cluster separation}

Given a solution $\sigma$,  a {\em flippable set} is a set of variables $S$ such that flipping the value of all variables in $S$ yields another solution $\tau$. Proving Theorem~\ref{t2} boils down to proving that {\whp}, in the subsystem induced by the 2-core,  every flippable set other than a flippable cycle has linear size.

A common approach to proving analogous statements is to establish that every flippable set, {other than a flippable cycle,} must deterministically induce a dense subgraph. In particular, if one can prove that for some constant $\e >0$, every such set is at least $1+ \e$ times as dense as a flippable cycle, then standard arguments yield the desired conclusion. Here, though, this is not the case, due to the possibility of arbitrarily long {paths of degree 2 vertices. Specifically, by replacing the edges of {any} flippable set (that is not a flippable cycle) by  {\em 2-linked paths}, one can easily create flippable sets whose density is arbitrarily close to that of a flippable cycle (for a more more precise statement, see the definition of {\em 2-linked paths} in Section \ref{spt2}).} Thus, controlling the number and interactions of these {2-linked} paths, an approach similar to that of~\cite{ABM,ms}, is crucial to our argument. In order to  work on the 2-core, we carry this analysis out on hypergraphs with a given degree sequence.

The key to controlling 2-linked paths is to bound a parameter governing the degree to which they tend to branch. Lemma \ref{lc*} shows that this {parameter} is bounded below 1, so while arbitrarily long 2-linked paths will occur, their frequency decreases exponentially with their length.

{We note that if we were working on hypergraphs} with minimum degree at least 3, then there would be no 2-linked paths, and the proof would have been very easy.  All of the difficulties arise from the problem of degree 2 vertices. {We note that our approach} applies to general degree sequences of minimum degree 2.

\subsection{Theorem~\ref{t1}: Connectivity inside clusters} \label{st1}
The main step in the proof of Theorem~\ref{t1} is to prove Theorem~\ref{tcore}; i.e. that every vertex outside
the $r$-core can be removed by an $r$-stripping sequence of length $O(\log n)$.

 It is often useful to consider stripping the vertices in several {parallel} rounds.

\begin{definition} The {\em parallel $r$-stripping process} consists of iteratively   removing {\em all} vertices of degree less than $r$ at once along with any hyperedges containing any of those vertices, until no vertices of degree less than $r$ remain.
\end{definition}

To prove that all non-core vertices can be removed by a stripping sequence of length $O(\log n)$, our approach is significantly different below and above the threshold, $c^*_{k,r}$, for the emergence of an $r$-core in random $k$-uniform hypergraphs. In both cases, we begin by stripping down to $H_B$, {the hypergraph remaining after $B$ rounds of the parallel stripping process,} for a sufficiently large constant $B$.  A simple argument shows that for any non-core vertex $v$, the number of vertices removed during {this initial phase that are} relevant to the removal of $v$, is bounded. {Thus,} what remains is to show that {any} non-core vertex in $H_B$ can be removed from $H_B$ by a stripping sequence of length $O(\log n)$.

For $c < c_{k,r}^*$, we prove that there exists a sufficiently large constant $B=B(c,k,r)$ such that all  connected components of $H_B$ have size at most $W = O(\log n)$; therefore, any remaining vertex can be removed with an additional $W$ strips. To do this we establish analytic expansions for the degree sequence of $H_B$ as $B$ grows and then apply a hypergraph extension of the main result of Molloy and Reed~\cite{mr} regarding the component sizes of a random $k$-uniform hypergraph with a given degree sequence.

For $c > c_{k,r}^*$, a lot more work is required. Once again, 2-linked paths are a major problem.
Indeed, it is not hard to see that a long 2-linked path with one endpoint of degree 1, can create a long stripping sequence leading to the removal of its other endpoint.

We first establish that for any $\e>0$, there exists a sufficiently large constant $B=B(c,k,r,\e)$ such that $H_B$ is sufficiently close to the $r$-core for two important properties to hold {in $H_B$}: (i) there are at most $\e n$ vertices of degree less than $r$, and (ii) the ``branching'' parameter for 2-linked paths, mentioned above, is bounded below 1.  Property (ii) allows us to control long 2-linked paths. However, this does not suffice as we need to control, more generally, for large tree-like stripping sequences. To do so, we note that any large tree must either have many leaves, or long paths of degree 2 vertices.  Such long paths will correspond to 2-linked paths in the random hypergraph, and so (ii) allows us to control the latter case.  Leaves of the tree will have degree less than $r$, and so (i) enables us to control the former case.

\section{{An Algorithm for Traveling Inside Clusters}}\label{sec:intra}
In this section, we show how we use Theorem~\ref{tcore} to prove Theorem~\ref{t1}.  In fact, we require
Theorem~\ref{main_lemma} below, which is  somewhat stronger than Theorem~\ref{tcore}.

Given a hypergraph $H$, we consider any terminal $r$-stripping sequence, $v_1,\ldots,v_t$, i.e., one that removes every vertex outside of the $r$-core of $H$.  Let $H_i$ denote the hypergraph remaining after removing $v_1,\ldots,v_{i-1}$; so $H_1=H$ and $H_{t+1}$ is the $r$-core of $H$.  Let $E_i$ denote the set of at most $r-1$ hyperedges in $H_i$ that contain $v_i$. We form a directed graph, $D$, as follows:
\begin{definition}
The vertices of $D$ are the non-$r$-core vertices $v_1,\ldots,v_t$, as well as any $r$-core vertex that shares a hyperedge with a vertex not in the $r$-core.  For each vertex $v_i$ in the stripping sequence, $D$ contains a directed arc $(u,v_i)$ for every vertex $u\neq v_i$ contained in the hyperedges of $E_i$.   Note that if $v_i$ has degree zero in $H_i$, then $E_i=\emptyset$, and so $v_i$ will have indegree zero in $D$.
\end{definition}

For every vertex $v$ in $D$, we define $R^+(v)$ to be the set of vertices that can be reached from $v$. Note that if $v$ is not in the $r$-core, then the vertices of $R^+(v)$ can be arranged into a (not necessarily terminal) $r$-stripping sequence ending with $v$.  So to prove Theorem~\ref{tcore}, it suffices to show $|R^+(v)|=O(\log n)$ for every such $v$.

\begin{theorem}\label{main_lemma}
For any integers  $k\geq 2,r\geq 2$ and any constant $c>0, c\neq c_{k,r}^*$, let $H = H_k(n,p=c/n^{k-1})$.   There exists a constant $Q=Q(k,r,c)>0$ such that \whp\ there is a {terminal} $r$-stripping sequence of $H$ for which in the digraph $D$ associated with the sequence:
\begin{enumerate}
\item[(a)] For every vertex $v$, $|R^+(v)| \leq Q\log n$.
\item[(b)] For $r=2$, for every {core}  flippable cycle $C$,
\[\sum_{v\in C}|R^+(v)| \leq Q\log n \enspace .\]
\end{enumerate}
\end{theorem}

\begin{remark}
{\em The proof of {Theorem}~\ref{main_lemma} can be extended to show that \whp\ for every vertex $v\in D$, the subgraph induced by $|R^+(v)|$ has at most as many arcs as vertices. }
\end{remark}
\begin{remark} 
{\em 
The case $r=k=2$ follows from previously known work.  For $c<c^*_{2,2}=1$, it follows from the fact that \whp\ every component of $G_{n,p=c/n}$ has size $O(\log n)$ below the giant component threshold $c^*_{2,2}$.  For $c>c_{k,r}^*$, it follows from Lemma 5(b) of \cite{pittel}.  Our proof will work for $k=r=2$, but it is convenient to assume $(k,r)\neq (2,2)$.}
\end{remark}
\begin{remark} {\em
We think that the conclusion of Theorem~\ref{main_lemma} does not hold at $c=c^*_{k,r}+o(1)$.  This is known to be true for the case $k=r=2$.  Indeed, when $c=1-\l$, for $\l=n^{-1/3+\e},\e>0$, \whp\  the size of the largest component
is $\Theta(\l^{-2}\log n)$ and no component has more than one cycle~\cite{tlcomp}.  A simple first moment analysis
yields that \whp\ there is no cycle of length greater than $\log n/\l$. Furthermore, \whp\ no vertex has degree greater than $\log n$.  It follows that the largest
component must contain an induced subtree, none of whose vertices are in the 2-core, which has size $\Theta(\frac{\l^{-2}\log n}{\log^2 n/\l})=\Theta(1/(\l\log n))$.  It is easy to see that such a subtree will contain vertices with depth $\Theta(1/(\l\log n))$, which can be as large as $n^{\a}$ for any $\a<1/3$.  }
\end{remark}

The proof of Theorem~\ref{main_lemma} occupies Sections~\ref{sec:above} and~\ref{sec:below}, after we set out some basic facts about cores in Section~\ref{sec:rgback} and some basic calculations in Section~\ref{sec:prelim}.
{But first, we show that it yields Theorems~\ref{t1} and~\ref{tcore}:}

\begin{proof}[Proof of Theorem~\ref{tcore}]  This follows immediately from Theorem~\ref{main_lemma} because the depth of $v$ is at most $|R^+(v)|$.
\end{proof}

{We are now ready to give our algorithm for traveling between any two assignments in the same cluster while changing $O(\log n)$ variables at a time.}
\begin{proof}[Proof of Theorem~\ref{t1}]
Given an arbitrary system of linear equations consider  a terminal $2$-stripping sequence $v_1,\ldots,v_t$ of its associated hypergraph and let $D$ be the digraph formed from the sequence.  For each core flippable cycle, $C$, we choose an arbitrary vertex $v_C\in C$. Let $B$ be the set consisting of each vertex $v_C$ and every non-2-core vertex with indegree zero in $D$.

Consider any 2-core solution $\s$.  Consider the system of equations formed from our system by fixing {the value of every 2-core variable that does not belong to a core flippable cycle to its value in $\sigma$;} we call such vertices \emph{fixed vertices}.  Recall from Definition \ref{flippable_def} that the edges of a flippable cycle contain vertices that are not considered to be vertices of the flippable cycle; such vertices will be fixed. Note that the solutions of this system form a cluster, and that every cluster can be formed in this way from some $\s$.

We will perform Gaussian elimination on this system in a manner such that $B$ will be the set of free variables that we obtain.  Importantly, this set of free variables does not depend on $\s$, i.e., it will be the same for every cluster.

For each $v\in B$ and for each fixed vertex $v$, set $\chi(v)=\{v\}$. For each core flippable cycle $C$, we process all of the edges (i.e., equations) joining consecutive vertices of $C$ except for one of the edges containing $v_C$. For each vertex $v\in C$, we obtain the equation $v=v_C+z_v$ where $z_v$ is a constant (0 or 1) depending only on  the assignment to the fixed vertices  in the edges of $C$; we set $\chi(v)=\{v_C\}$.   By Lemma~\ref{flip_cycles}, the core flippable cycles are vertex-disjoint, and so the equations corresponding to two core flippable cycles can overlap only on fixed variables.
Thus,  we can carry this out for each core flippable cycle $C$ independently.

Next,  we  process the edges not in the 2-core, in reverse removal order, i.e., $E_t,\ldots,E_1$.  Note that, since $r=2$, each $E_i$ contains at most one edge. When processing $E_i$, we set  $\chi(v_i)$ to  the {symmetric difference} of the sets $\chi(u)$, over all $u\in E_i$ other than $v_i$. That is, a variable $z$ is in $\chi(v_i)$ iff $z \in\chi(u)$ for an odd number of variables $u\in E_i$ other than $v_i$. Since $E_i$ is the equation $v_i=\sum_{u\in E_i; u\neq v}u$, this is equivalent (by induction) to $v_i=\sum_{w\in \chi(v_i)}w+z_{v_i}$, where $z_{v_i}$ is the sum of $z_u$ over all vertices $u\in \chi(v_i)$ that belong to core flippable cycles. We now note that every non-2-core  vertex $v_i\notin B$ has indegree at least 1 in $D$ and so $|E_i|=1$ and thus $\chi(v_i)$ is defined.  For each vertex $u\neq v_i$ in $E_i$, either $u\in B$, or $u$ is fixed, or $u=v_j$ for some $j>i$, or $u$ is in a core flippable cycle. Therefore, by induction, $\chi(v_i)$ contains only vertices that are  in $B$ or are fixed.

Finally, note that possibly $\chi(v_i)=\emptyset$; in that case, $v_i=\sum_{w\in \chi(v_i)}w+z_{v_i}=z_{v_i}$ in every solution.  (It is not hard to adapt the proof of Theorem~\ref{main_lemma} to show that \whp\ for every $i$, $\chi(v_i)\neq\emptyset$.  But that is not required for the purposes of this paper.)

At this point, all non-fixed vertices are either in $B$ or have been expressed as the sum of vertices in $B$ and fixed vertices. Therefore, the vertices in $B$ are the free variables for the system obtained by fixing the values of the fixed vertices to $\s$. Thus, there are exactly $2^{|B|}$ solutions to that system, one for each assignment to $B$.  We can move between any two such solutions by changing the assignments to the vertices of $B$, one at a time.  Each time we change the value of a non-2-core vertex $v \in B$, in order to get to another solution, we only need to change a subset of $R^+(v)$ in the digraph $D$, because only vertices $u\in R^+(v)$ can have $v\in\chi(u)$.  Similarly, each time we change the value of some $v_C\in B$, we only need to change a subset of $\cup_{v\in C}R^+(v)$. Thus, by Theorem~\ref{main_lemma}, we can move between any two such solutions changing at most $Q\log n$ variables at a time. This implies Theorem \ref{t1}, since each cluster is such a solution set.
\end{proof}

We close this section by showing how the preceding proof extends to determine all of the frozen variables. A variable is said to be \emph{frozen} in a cluster, if it takes the same value in all  assignments of the cluster. In general {random CSPs} it is hypothesized that the set of frozen variables can differ from cluster to cluster. In random $k$-XOR-SAT, though, the set of frozen variables depends only on the underlying hypergraph, i.e., is the same for all clusters.


\begin{theorem}\label{tf} In every cluster, the frozen variables consist of the 2-core vertices not in {core} flippable cycles, and the non-2-core variables $v$ for which $\chi(v)\cap B=\emptyset$.
\end{theorem}

\begin{proof}  This follows immediately from the fact that $B$ is the set of free variables
in a system of linear equations whose solution set is the cluster.
\end{proof}

\section{Random hypergraphs and their cores}\label{sec:rgback}
We will use the configuration model of Bollob\'as~\cite{bb} to generate a random $k$-uniform hypergraph $H$ with a given degree sequence.  Suppose we are given the degree $d(v)$ for each vertex $v$; thus $\sum d(v)=kE$ where $E$ is the number of hyperedges. We take $d(v)$ {\em copies} of each $v$, and we take a uniformly random partition of these $kE$ vertex-copies into $E$ sets of size $k$. This naturally yields a $k$-uniform hypergraph,  by mapping each $k$-set to a hyperedge on the vertices whose copies are in the $k$-set. Note that the hypergraph may contain loops (two copies of the same vertex in one hyperedge) and multiple edges (two identical hyperedges). It is well known that the probability that this partition yields a simple hypergraph (i.e., one with no loops or multiple edges) is bounded below by a constant for degree sequences\footnote{Clearly, we are referring to a sequence of degree sequences $\mathcal{S}_n$ so that asymptotic statements are meaningful. We suppress this point though, throughout, to streamline exposition.} satisfying certain conditions. {Specifically:}
\begin{definition}
Say that a {degree sequence $\mathcal{S}$} is \emph{nice} if $E=\Theta(n)$, $\sum_v d(v)^2=O(n)$
and $d(v)=o(n^{1/24})$ for all $v$.
\end{definition}

Every degree sequence we will consider will correspond to some subgraph of $H_k(n,p)$ with a linear {expected} number  of edges. Since, as is well known, the degree sequence of {such random hypergraphs} is nice \whp, all the degree sequences we will consider will be nice. With this in mind, we will make heavy use of the following standard proposition (see eg. \cite{cc}) and corollary, as working in the configuration model is technically much easier than working with uniformly random  hypergraphs with a given degree sequence.

\begin{proposition} If  $\mathcal{S}$ is {a nice degree sequence,} then there exists $\e>0$ such that the probability that
a random hypergraph with degree sequence $\mathcal{S}$ {{drawn from} the configuration model} is simple is at least $\e$.
\end{proposition}

This immediately yields:
\begin{corollary}\label{ccon}
If  $\mathcal{S}$ is a nice degree sequence then:
\begin{enumerate}
\item[(a)] If property $Q$ holds \whp\ for $k$-uniform hypergraphs with degree sequence $\mathcal{S}$ {drawn from} the configuration model, then $Q$ holds \whp\ for uniformly random simple hypergraphs with degree sequence $\mathcal{S}$.
\item[(b)]
For any {random} variable $X$, if $E(X)=O(1)$ for $k$-uniform hypergraphs with degree sequence $\mathcal{S}$ {drawn from} the configuration model, then $E(X)=O(1)$ for uniformly random simple hypergraphs with degree sequence $\mathcal{S}$.
\end{enumerate}
\end{corollary}

The following lemma will be very useful. Its exponential term is not tight, but will suffice for our purposes.

\begin{lemma}\label{lconfig}
Consider a random $k$-uniform {hypergraph drawn from the configuration model} with $E$ edges, i.e., with total degree $kE$.  For each $i=2,\ldots,k$, specify $\ell_i$ sets of $i$ vertex-copies, and
set $L=\sum_{i=2}^k\ell_i$. The probability that each of these sets appears in some hyperedge, and no two appear in the same hyperedge is less than
\[
\exp\left(\frac{kL^2}{E-L}\right)
\prod_{i=2}^k\left(\frac{(k-1)(k-2)\cdots(k-i+1)}{(kE)^{i-1}}\right)^{\ell_i} \enspace .\]
\end{lemma}
\begin{proof}

We choose the {partition of the vertex-copies}  by processing the specified sets one-at-a-time. To process one of the $\ell_i$ sets of size $i$, we first choose one set member $\g$ arbitrarily and then randomly select the remaining $k-1$ vertex-copies of the {part} containing $\g$. Every time we do this there are at least $kE-kL$ yet unselected vertex-copies. Thus, the probability we chose all other $i-1$ members of the specified set is at most
\begin{eqnarray*}
\frac{(k-1)(k-2)\cdots(k-i+1)}{(kE-kL)^{i-1}} &<&\frac{(k-1)(k-2)\cdots(k-i+1)}{(kE)^{i-1}} \times \left(\frac{E}{E-L}\right)^{i-1}\\
&<&\frac{(k-1)(k-2)\cdots(k-i+1)}{(kE)^{i-1}} \times e^{kL/(E-L)},
\end{eqnarray*}
since $i\leq k$. So the probability that {each of the} $L$ tuples {is} chosen to be in a hyperedge is less than
\[\prod_{i=2}^k\left(\frac{(k-1)(k-2)\cdots(k-i+1)}{(kE)^{i-1}}\right)^{\ell_i}\times e^{(kL/(E-L))\ell_i}
=e^{kL^2/(E-L)}\times\prod_{i=2}^k\left(\frac{(k-1)(k-2)\cdots(k-i+1)}{(kE)^{i-1}}\right)^{\ell_i}
\enspace .
\]

\end{proof}

\subsection{Cores}\label{sc}

Recall from Section \ref{sec:intra} that Theorem~\ref{main_lemma} is already known for $k=r=2$.
So we will assume that {$k+r > 4$.} It is well known that the $r$-core {of a random $k$-uniform hypergraph} is {uniformly random conditional on} its degree sequence. See \cite{psw} for the case $k=2$, and \cite{mmcore} for the nearly identical proof for general $k$.  In fact, the same is true of the graph remaining after any number of iterations of the parallel stripping process.

Let {$H=H_k(n,p)$} be a random $k$-uniform hypergraph and let ${H}=H_0, H_1, \ldots$ be the sequence of hypergraphs produced by the parallel {$r$-stripping} process.
It is {well known how} (see e.g., \cite{mmcore}) to show the following propositions.
\begin{proposition}\label{pdeg}
\mbox{}
\begin{enumerate}
\item[(a)] For every $i\ge 0$, $H_i$ is uniformly random with respect to its degree sequence.
\item[(b)] There exist functions $\r_0,\r_1,\ldots$ such that for any {fixed integer $i$, \whp\ $H_i$} contains ${\r_j(i)}n+o(n)$ vertices of degree $j$ and $\frac{1}{k}(\sum_{j\geq 1} j {\r_j(i)})n+o(n)$ edges.
\end{enumerate}
\end{proposition}

\begin{remark}\emph{
The functions $\r_j(i)$ have explicit recursive expressions, which we give in Section \ref{sec:below}.  An approximation is stated in Proposition \ref{lT} below.}
\end{remark}

Proposition \ref{pdeg} allows us to use the configuration model to study $H_i$.  We will begin by showing that we can uniformly approximate the total degree of $H_i$.
\begin{lemma}\label{ltail}  
For every {fixed integer} $i\geq 0$,
\[
\sum_{v\in H_i}\deg_{H_i}(v)=\left(\sum_{j\geq 1}j\r_j(i)\right)n +o(n) \enspace .
\]
\end{lemma}

\begin{proof}
Proposition \ref{pdeg} implies that $\sum_{j\geq 1}j\r_j(i)$ is convergent, else \whp\ $H_i$, and hence $H$, would have a superlinear number of edges.

Consider any fixed $J$. By Proposition \ref{pdeg}, \whp\ $\sum_{v:\deg_{H_i}(v)\leq J}\deg_{H_i}(v)=\sum_{j= 1}^Jj\r_j(i)n+o(n)$.  For any $\theta>0$, the convergence of $\sum_{j\geq 1}j\r_j(i)$ implies that we can
choose $J=J(\theta)$ sufficiently large that $\sum_{j>J}j\r_j(i)<\theta/2$.   Since $H_i\subseteq H_0 =H$, we have $\sum_{v:\deg_{H_i}(v)>J}\deg_{H_i}(v)\leq\sum_{v:\deg_{H}(v)>J}\deg_{H}(v)$. The fact that the latter sum is less than $\theta n/2$ for $J$ sufficiently large is well known and follows from the facts that (i) for each constant $\ell$, the number of vertices of degree $\ell$ in $H$ is \whp\  $\l_{\ell}n+o(n)$ for a particular $\l_{\ell}=\l_{\ell}(c)$ and (ii) the number of hyperedges in $H$ is highly concentrated around $\inv{k}\sum_{\ell\geq1} \ell\l_{\ell}n$. Thus, 
$\Bigl|\sum_{v\in H_i}\deg_{H_i}(v)-\left(\sum_{j\geq 1}j\r_j(i)\right)n\Bigr|<\theta n$ for every $\theta>0$, which establishes the lemma.
\end{proof}

The following similar bound will also be useful:

\begin{lemma}\label{ltail2}  For every constant $d$ and fixed {integer} $i>0$:
\[
\sum_{v:\deg_{H_i}(v)\geq d}\frac{\deg_{H_i}(v)!}{(\deg_{H_i}(v)-d)!}=\left(\sum_{j\geq d}\frac{j!}{(j-d)!}\r_j(i) \right)n+o(n) \enspace .
\]
\end{lemma}
\begin{proof}
The proof is almost identical to that of Lemma \ref{ltail} but exploits the concentration of the number of $d$-stars in $H$, rather than of the number of hyperedges. (A $d$-star is a set of $d$ hyperedges which contain a common vertex.) The concentration of the number of $d$-stars in $H$ is easily established, e.g., by the Second Moment Method or Talagrand's Inequality. (Indeed, Lemma \ref{ltail} and its proof are special cases of this lemma and its proof for $d=1$.)
\end{proof}

For any fixed integers $k,r$ and real number $\lambda > 0$, we write
$$
\Psi_r(\lambda) = e^{-\lambda}\sum_{i \ge r-1} \l^i/i! \qquad \mbox{and} \qquad
f_{k,r}(\lambda) = f(\l)=\frac{(k-1)!\l}{\Psi_{r}(\lambda)^{k-1}} \enspace .
$$
Recall  that for $k+r>4$, the threshold for the appearance of an $r$-core in a random $k$-uniform hypergraph $H_k(n,p)$ with $p=c/n^{k-1}$  is
\[c^*_{k,r}=\min_{\l>0} f_{k,r}(\lambda).\] We will see that $f'$ has a unique root and, thus,  for $c > c^*_{k,r}$ the equation
$f(\l) = c$ has two solutions.
\begin{definition}\label{mu_def}
For $c > c^*_{k,r}$, let $\mu=\mu(c)$ denote the larger of the two solutions of $f(\l) = c$.
\end{definition}
The following two propositions are standard; see {e.g.,}~\cite{mmcore} for proofs.

\begin{proposition}\label{ldeg} For every fixed $j\geq r$, \whp\ the $r$-core contains $(e^{-\mu} \mu^j/j!)n +o(n)$ vertices of degree $j$. Furthermore, \whp\ the $r$-core contains
$(\mu/k) \Psi_r(\mu)n+o(n)$ edges.
\end{proposition}

\begin{proposition}\label{lT} For every $c\neq c_{k,r}^*$ and  $\theta>0$, there exists $B = B(\theta)$ such that \whp\
\begin{enumerate}
\item[(a)] $H_B$ contains fewer than $\theta n$ vertices not in the $r$-core;
\item[(b)] For each $j\geq r$, $|\r_j(B) - e^{-\mu} \mu^j / j!|<\theta$.
\end{enumerate}
\end{proposition}

{The following lemma will be critical for our analysis.}

\begin{lemma}\label{lc*}
For every $c> c^*_{k,r}$, there exists $\z = \z(k,r,c)>0$ such that
\begin{equation}\label{eq:critical}
(k-1) \frac{\mu^{r-1}}{(r-2)!}<(1-\z)\sum_{i\geq r-1}\frac{\mu^i}{i!} \enspace ,
\end{equation}
where $\mu$ is the {larger} of the two roots of the equation $f_{k,r}(\lambda)=c$.
\end{lemma}
\begin{proof}
\begin{equation}\label{eq:oneline}
f'(\l) = 0
\quad \Longleftrightarrow \quad
\Psi_r(\l) =
\l (k-1) \Psi_r(\l)^{k-2} \Psi'_r(\l)
\quad \Longleftrightarrow \quad
\sum_{i\geq r-1}\frac{\l^i}{i!}  =
(k-1)  \frac{ \l^{r-1}}{(r-2)!} \enspace .
\end{equation}
Equation~\eqref{eq:oneline} yields $c^*_{k,r}=f(\l^*)$ for some $\l^*$ satisfying the last equation in~\eqref{eq:oneline}.
{For $c>c^*_{k,r}$, since $\mu = \mu(c)$ is the {larger} of the two roots of $f(\l) = c$, it follows that $\mu>\l^*$.} The lemma now follows by noting that {the RHS of~\eqref{eq:critical} divided by the LHS} is proportional to $\sum_{i\geq r-1}\frac{\mu^{i-r+1}}{i!}$, which is clearly increasing with $\mu$.
\end{proof}



\section{Preliminaries to the proof of Theorem~\ref{main_lemma}}\label{sec:prelim}

Recall that we assume $k+r>4$ and let $H=H_k(n,p)$ be a random $k$-uniform hypergraph with $p=c/n^{k-1}$. Let $H=H_0,H_1,\ldots$ be the sequence of hypergraphs produced by the parallel $r$-stripping process.

As we said above, we will choose a sufficiently large constant $B$,  strip down to $H_B$, and then focus on  $R^+(u)\cap H_B$, making use of the fact that $H_B$ is very close to the 2-core (by Proposition \ref{lT}).
The following will be used to bound the number of vertices that are removed from $R^+(u)$ when stripping down to $H_B$. For integer $s\geq 0$, we use $N^s(v)$ to denote the $s$-th neighborhood of $v$, i.e., the set of vertices within distance $s$ from $v$. For any set of vertices $A$,  $N^s(A)=\bigcup_{v\in A}N^s(v)$. We consider a single vertex {to be} a connected set. A straightforward induction yields the following.

\begin{proposition}\label{pns}
For any integer $i$ and vertex $u\in H_i$,  $R^+(u)\subseteq N^i(R^+(u)\cap H_i)$.
\end{proposition}

\begin{lemma}\label{lst}
For any $c,s \ge 0$, there exists $\Gamma=\Gamma(c,s)$ such that in a random graph $G(n,p)$ with $p=c/n$, \whp\ for every connected subset $A$ of vertices $|N^s(A)|\leq \Gamma(|A|+\log n)$.
\end{lemma}

\begin{proof}
We prove this for the case $s=1$, i.e., that there is a constant $\gamma>1$ such that \whp\ every connected subset of vertices $A$ satisfies
$|N(A)|\leq \gamma(|A|+\log n)$. By iterating, we obtain that for every $s \ge 1$, every connected subset of vertices $A$ satisfies $|N^s(A)|\leq f_s(|A|)$
where
\begin{eqnarray*}
 f_1(x)&=&\gamma(x+\log n)\\
f_{i+1}(x)&=&\gamma(f_i(x)+\log n) ,  \mbox{ for $i\geq 1$.}
\end{eqnarray*}
A simple induction yields $f_{i}(x)\leq \gamma^i(x+ i\log n)$ and that yields the lemma with $\Gamma = s \gamma^s$.

Given any set $A$ of size $a$, the probability that $A$ is connected is at most the expected number of spanning trees of $A$ which is $a^{a-2}(c/n)^{a-1}$.  After conditioning that $A$ is connected, the number of neighbors outside of $A$ is distributed as ${\mathrm{Bin}}(a(n-a),c/n)$. The probability that this exceeds $z$ is at most
\[\binom{a(n-a)}{z}\left(\frac{c}{n}\right)^z<\left(\frac{eca}{z}\right)^z<2^{-z},\qquad\mbox{for $z>2eca$.}\]
 For any $\gamma>2$, if $|N(A)|> \gamma(|A|+\log n)$, then we must have $|N(A)\bk A|> \hf \gamma(|A|+\log n)$. Taking $\gamma>4ec$, the expected number of connected sets $A$ satisfying this last inequality is at most
\[{n\choose a}a^{a-2}\left(\frac{c}{n}\right)^{a-1}2^{-\hf \gamma(a+\log n)}
<\frac{en}{a^2}\left(ec\right)^{a-1}2^{-\hf \gamma(a+\log n)}
<\frac{en}{a^2}\left(\frac{ec}{2^{\gamma/2}}\right)^{a-1}2^{-\hf \gamma\log n}=n^{-\Theta(\gamma)}\enspace ,\]
for $\gamma$ sufficiently large.  Multiplying by the $n$ choices for $a$ yields the lemma.
\end{proof}

\begin{lemma}\label{flip_cycles}
Fix $k \ge 3$ and let $H=H_k(n,p)$ be a  random $k$-uniform hypergraph with  $p = c/n^{k-1}$, where $c>c^*_{k,2}$. 
\begin{enumerate}
\item[(a)] The expected  number of vertices in {core} flippable cycles  of $H$ is $O(1)$.
\item[(b)] W.h.p.\ no vertex lies in two core flippable cycles.
\end{enumerate}
\end{lemma}

\begin{proof}
Let $\cald$ be the degree sequence of the 2-core of $H$. By  Corollary \ref{ccon},  we can work in the configuration model.
Recalling Definition~\ref{mu_def}, Proposition~\ref{ldeg} and Lemma \ref{lc*}, \whp\
\begin{enumerate}
\item[(i)] $\cald$ has total degree $\g n+o(n)$, where $\gamma = \mu \Psi_r(\mu)$,
\item[(ii)] $\cald$ has $\la_2 n+o(n)$ vertices of degree 2,  where $\lambda_2 = e^{-\mu} \mu^2/2$,
\item[(iii)]  there exists $\z >0$ such that
$2(k-1)\la_2<(1-\z)\g$.
\end{enumerate}

We first bound
the expected number of {core} flippable cycles of size $a$. Let $\Lambda=\g n+o(n)$ be the total number of vertex copies, and let $L=\la_2 n+o(n)$ be the number of copies of degree 2 vertices.

There are ${L\choose a}$ choices
for the connecting vertices, $\frac{(a-1)!}{2}$ ways to order them into a cycle, and $2^a$ ways to align their vertex-copies.
This yields $a$ pairs $\{y_1,z_1\},\ldots,\{y_a,z_a\}$ of vertex copies, each of which must land in a hyperedge. We process these pairs one-at-a-time, halting if we ever find that the pair does not land in a hyperedge. To process pair $i$, we ask only whether $z_i$ lands in the same hyperedge as $y_i$; if it does we do {\em not} expose the other vertex-copies in that hyperedge.  Thus, prior to processing pair $i$, we have exposed exactly $2i-2$ vertex-copies, all of degree 2.  There are $k-1$ other copies appearing in the same hyperedge as $y_i$. Each of the $\Lambda-(2i-1)$ unexposed copies (not including $y_i$) is equally likely to be one of those copies (and, for $k\geq 3$, the exposed copies also have positive probability).  So the probability that $z_i$ is one of them is at most $(k-1)/(\Lambda-2i+1)$.  So the expected number of {core} flippable cycles of length $a$ is at most:

\[{L\choose a}\frac{(a-1)!}{2}2^a\prod_{i=1}^a\frac{k-1}{\Lambda-2i+1}
<\inv{2a}\prod_{i=1}^a\frac{2(k-1)(L-i+1)}{\Lambda-2i+1}.\]
By condition (iii) above, $2(k-1)L/(\Lambda-1)<1-\hf\z$, and so $2(k-1)(L-i+1)/(\Lambda-2i+1)<1-\hf\z$ for each $i$, since $L \le \hf(\Lambda-1)$. So the expected number is at most $\inv{2a}(1-\hf\z)^a$, and so the expected total number of vertices on {core} flippable cycles is at most $\hf\sum_{a\geq 1} (1-\hf\z)^a=O(1)$. This establishes part (a).

We now prove part (b) by using a first moment calculation.  We start by showing that if two core flippable cycles have a common vertex, then their union must contain a simple structure: a flippable cycle plus a path.   The proof will then follow since \whp\ any such structure in a sparse random hypergraph must be larger than what is permitted by part (a).  Intuitively, this is straightforward, but the details are tedious.  

Recall that for every flippable cycle $C$ of a hypergraph $H$, every vertex of $C$ has degree two in $H$ and every edge of $H$ contains either zero or exactly two vertices of $C$. Let $S = x_1,x_2,\ldots,x_a$ be a core flippable cycle and let $S' \neq S$ be any other core flippable cycle that contains $x_1$, where without loss of generality we assume $|S'| \le |S|$. Let $e_1$ be the edge incident to $S$ that contains $x_1, x_2$  and let $e_a$ be the edge incident to $S$ that contains $x_1,x_a$. Since $e_1,e_a$ are the only two edges incident to $x_1$, they must be among the edges incident to $S'$. Let $y_2 \neq x_1$ be the other vertex of $S'$ in $e_1$. If $y_2 = x_2$, let $e_2$ be the edge containing $x_2,x_3$ and let $y_3 \neq y_2$ be the other vertex of  $S'$ in $e_2$. If $y_3 = x_3$, let $e_3$ be the edge containing  $x_3,$ etc.\ until we first reach an edge $e_j$ containing $x_j, x_{j+1}$ such that $y_{j+1} \neq x_{j+1}$. Let $e'_{j+1}$ be the edge containing $y_{j+1}, y_{j+2}$. Now, let $e'_{j+2}$ be the edge containing $y_{j+2}, y_{j+3}$, let $e'_{j+3}$ be the edge containing $y_{j+3}, y_{j+4}$ etc., until we first reach an edge $e'_{j+1+q}$ containing a vertex lying in a hyperedge incident to $S$ (this must occur for some $q \ge 0$ since $e_a$ is incident to $S'$). Thus, we have a path $y_{j+1}, \ldots,y_{j+1+q+1}$ where $y_{j+1}$ and $y_{j+1+q+1}$ lie in hyperedges incident to $S$, while all other $q$ vertices $y_{j+2}, \ldots, y_{j+1+q}$ do not lie in hyperedges incident to $S$. In the following, to lighten notation, we let $j+1 = \ell$ and $j+1+q+1 = \ell^*$, i.e., $y_{\ell}$ and $y_{\ell^*}$ are the vertices in the path lying in hyperedges incident to $S$.

Now we bound the expected number of occurences of such a  cycle $S$ of size $a$ plus a path $y_{\londara},...,y_{\kwuumba}$, in $H=H_k(n,p)$. Since $|S|=a$ there are at most $a^2$ choices for the value of ${\londara}$ and the value of $z$ such that $x_{z-1},x_{z},y_{\londara}$ share a hyperedge. Since $|S'| \le |S|$ there are at most $a$ choices for the value of $\ell^*$. Setting $b=\kwuumba-{\londara}+1$ to be the number of vertices in the path, the number of choices for $x_1,...,x_a,y_{\londara},...,y_{\kwuumba}$ is at most $n^{a+b}$.  The probability that $x_{{\londara}-1},x_{\londara},y_{\londara}$ share a hyperedge in $H$ is at most ${n\choose k-3}\frac{c}{n^{k-1}}$, and the same is true for $x_{z-1},x_{z},y_{\kwuumba}$. For each $i\neq {\londara},z$, the probability that $x_{i-1},x_{i}$ share a hyperedge is at most ${n\choose k-2}\frac{c}{n^{k-1}}$.  For each $i= {\londara},...,\kwuumba-1$, the probability that $y_i,y_{i+1}$ share a hyperedge is at most ${n\choose k-2}\frac{c}{n^{k-1}}$.  So the expected number of such subgraphs is at most:
\[a^3 n^{a+b}\left({n\choose k-3}\frac{c}{n^{k-1}}\right)^2\left({n\choose k-2}\frac{c}{n^{k-1}}\right)^{a+b-3}<\frac{a^3c^{a+b-1}}{n}=o(1),\]
for $a+b\leq \hf\log n$.  Thus \whp\ there is no such subgraph with $a+b\leq \hf\log n$.  Since $|S|\geq |S'|$, if $a+b> \hf\log n$ then $|S|=a>\inv{4}\log n$.   But part (a) and Markov's Inequality imply that \whp\ there is no core flippable cycle of size at least $\inv{4}\log n$.
This proves (b).

{\bf Remark:} By carrying out the first moment calculation more carefully, as in part (a), one obtains that the sum over all $a,b$ of the expected number  of occurrences of a core flippable cycle $S$ plus a path $y_{\londara},...,y_{\kwuumba}$ as described is, in fact, $O(n^{-1})$.

\end{proof}

\section{Proof of Theorem~\ref{main_lemma} above the $r$-core threshold}\label{sec:above}

Recall that we can assume $k+r>4$. We let $H=H_k(n,p)$ be a random $k$-uniform hypergraph with $p=c/n^{k-1}$. Let $H=H_0,H_1,\ldots$ be the sequence of hypergraphs produced by the parallel $r$-stripping process. We will choose a {terminal $r$-}stripping sequence that is consistent with the parallel process; i.e., in our stripping sequence: for every $i<j$, the vertices deleted in round $i$ of the parallel process come before  the vertices deleted in round $j$ of the parallel process.

Let $D$ be the digraph associated with this {terminal} $r$-stripping sequence and recall  that $R^+(u)$ denotes the set of vertices reachable from a vertex $u$ in $D$.

\subsection{{Bound on the length of stripping sequences}}\label{s5.1a}
Our main challenge is to prove the following lemma. The idea is that we will take {$B$ large enough} so that by stripping down to $H_B$, Proposition~\ref{lT} gives us control of the degree sequence that remains, and Lemma~\ref{lc*} allows us to prove that a certain branching process involving long paths in a graph constructed from $H_B$ dies out.
\begin{lemma}\label{near_core}
For every $c > c_{k,r}^*$ there exists $B=B(c,k,r)$ and $Q = Q(c,k,r)$ such that \whp\  for every  vertex $u$, $|R^+(u)\cap H_B| \leq Q\log n$.
\end{lemma}

\begin{proof}[Proof of Theorem~\ref{main_lemma}(a)]
Consider any vertex $u$. If $u\notin H_B$, then by Proposition \ref{pns}, $R^+(u)\subseteq N^B(u)$ in which case Lemma \ref{lst} immediately implies that $|R^+(u)|<\Gamma (1+\log n)$ for some constant $\Gamma = \Gamma(c,B)$.

If $u\in H_B$, then $R^+(u)\subseteq N^B(R^+(u) \cap H_B)$, by Proposition \ref{pns}. Since, by Lemma~\ref{near_core}, $|R^+(u)\cap H_B|\leq Q\log n$, Lemma~\ref{lst} now implies that $|R^+(u)|< \Gamma (Q\log n  + \log n) = Z \log n$ for $Z = \Gamma Q+1 = Z(c,B)=Z(c,k,r)$.
\end{proof}

\begin{definition} For any $i$, we define $D_i$ to be the subdigraph of $D$ induced by the vertices in $H_i$.
\end{definition}

Consider a particular constant $i$. Let $T^+$ be a directed tree in $D_i$ with edges directed away from a root $u$ that spans the vertices of $R^+(u)\cap H_i$; e.g., $T^+$ could be a {Breadth First Search} or {Depth First Search} tree from $u$. Thus, each vertex has indegree at most 1 in $T^+$, implying:
\begin{proposition}\label{p2arc}
No two arcs of $T^+$ were formed during the removal of the same hyperedge.
\end{proposition}
\begin{definition} A {\em deletion tree rooted at $u$} is the undirected tree, $T$, formed by removing the directions from a tree $T^+$ {rooted at $u$.}
\end{definition} 

To prove Lemma \ref{near_core}, we will bound the expected number of deletion trees $T$ of size greater than $Q\log n$. The following technical lemma bounds the density of small subgraphs of $H_k(n,p)$. It is of a standard flavour and has a standard proof. Given a subset $S$ of the vertices of $H_k(n,p)$, we let $\ell_j(S)$ denote the number of hyperedges
that contain exactly $j$ of the vertices of $S$, and we let $L(S)=\sum_{j=2}^k (j-1)\ell_j$.

\begin{lemma}\label{l22} For every $c,\z>0$, there is $\theta>0$, such that \whp\ every  $S\subseteq H_k(n,p=c/n^{k-1})$ with $|S|\leq\theta n$
has $L(S)<(1+\z)|S|$.
\end{lemma}

\begin{proof} Rather than working in the $H_k(n,p)$ model, it will be convenient to work in the $H_k(n,m)$ model, where exactly $m=(c/k!) n$ edges are selected uniformly, independently and with replacement (note that $m=p{n\choose k}$). Standard arguments imply that high probability properties in this model transfer to the $H_k(n,p)$ model.

Let $Y_a = Y_a(\z)$ denote the number of sets $S$ with $|S|=a$ and $L(S) = (1+\z)|S|$.  We will bound $\ex(Y_a)$ as follows.
Define
\[
\call_a=\left\{
(\ell_2,\ldots,\ell_k) :  \sum_{j=2}^k (j-1)\ell_j\geq (1+\z) a
\right\} \enspace .\]
Choose $a$ vertices and some $(\ell_2,\ldots,\ell_k)\in\call_a$,  pick $\ell_j$ edges for each $j$, and then multiply
by the probability that each edge chooses (at least) the appropriate number of vertices from $S$.
This yields
\begin{eqnarray*}
E(Y_a)& \le & {n\choose a}\sum_{(\ell_2,\ldots,\ell_k)\in\call_a}\prod_{j=2}^k
{m\choose\ell_j}\left[{k\choose j}\left(\frac{a}{n}\right)^j\right]^{\ell_j}\\
&<&\left(\frac{en}{a}\right)^a
\sum_{(\ell_2,\ldots,\ell_k)\in\call_a}\left(\frac{a}{n}\right)^{\sum_{j=2}^k(j-1)\ell_j}
\prod_{j=2}^k\frac{({J}a)^{\ell_j}}{\ell_j!} \quad , \quad \mbox{for some constant } {J =J(c,k)} > 0\\
&<&\left(\frac{en}{a}\right)^a\left(\frac{a}{n}\right)^{(1+\z)a}
\prod_{j=2}^k\left(\sum_{\ell_j\geq 0}\frac{({J}a)^{\ell_j}}{\ell_j!}\right)\\
&<&e^a\left(\frac{a}{n}\right)^{\z a}e^{(k-1){J}a}\\
&=&\left(\frac{\D a}{n}\right)^{\z a}  \quad , \quad \mbox{for some constant } \D=\D(c,k,\z) >0 \enspace .\\
\end{eqnarray*}
Choosing $\theta=\frac{1}{2\D}$, it is  standard and straightforward to show $\ex\left(\sum_{a=1}^{\theta n}Y_a\right)=o(1)$.
\end{proof}

In order to carry out our first moment calculation, we will bound the difference between the degrees of the vertices of $T$ and their degrees in $H_i$.

\begin{lemma}\label{pt} For any $\delta > 0$, if $i$ is sufficiently large in terms of $\d$ then \whp :
For every vertex $u\in D_i$, if $T$ is a deletion tree rooted at $u$, then
$\deg_{H_i}(v)\leq\deg_T(v)+r-2$ for all but at most $\d|T|+3$ vertices $v\in T$.

\end{lemma}

\begin{proof} 
Define $S$ to be the hypergraph with edge set
 $\{e\cap R^+(u):e\in H_i, |e\cap R^+(u)|\geq 2\}$.  In other words, for each hyperedge $e\in H_i$ that contains at least two vertices of $R^+(u)$,
 $S$ contains the edge obtained by removing all vertices outside of $R^+(u)$ from $e$.
 
Since $V(T)=R^+(u)\cap H_i$, the $r$-stripping sequence that yields $D$ contains an $r$-stripping subsequence which removes from $H_i$
only vertices of $T$, such that all vertices of $T$ except possibly $u$ are removed.
Consider $v\in T, v\neq u$. At the point that $v$ is removed, it has degree at most $r-1$ in what remains of $H_i$.  Every other hyperedge of $H_i$ containing $v$ is removed before $v$, and thus must contain another member of $R^+(u)$. At least one of those $r-1$ hyperedges contains another vertex of $R^+(u)$, namely the parent of $v$ in $T$. Therefore:
\[\deg_{H_i}(v)\leq \deg_{S}(v) +r-2.\]

For $2\leq j\leq k$, let $\ell_j$ denote the number of hyperedges with $j$ vertices in $S$.  All
vertices of $S$, except possibly $u$, are not in the $r$-core. So, by Lemma \ref{lT}(a), we know that for any $\theta>0$ we can select $i$ sufficiently large {in terms of $\theta$  so that} $|S|<\theta n$.  If we pick $\theta$ sufficiently small in terms of $\d$, then Lemma \ref{l22} implies that \whp,  $\sum_{j=2}^k (j-1)\ell_j <(1+\d/2)|S|$. So
\[\sum_{v\in R^+(u)}\deg_{S}(v)=\sum_{j=2}^k j\ell_j\leq2\sum_{j=2}^k (j-1)\ell_j <(2+\d)|S|=(2+\d)|T|.\]
Now the total $T$-degree of the vertices in $R^+(u)$ is $2|T|-2$, since $T$ is a tree with edges of size 2 that spans $R^+(u)$. So for  $i$ sufficiently large in terms of $\d$,
\[\sum_{v\in R^+(u)}\deg_{S}(v)-\deg_T(v)\leq(2+\d)|T|-(2|T|-2)=\d|T|+2.\]
So $\deg_T(v)\neq\deg_S(v)$ for at most $\d|T|+2$ vertices $v\in R^+(u)$. Also, $\deg_{H_i}(v)\leq \deg_{S}(v) +r-2$ for all but at most one  $v\in R^+(u)$ (namely $v=u$).  This proves the lemma.
\end{proof}


{\em Proof of Lemma ~\ref{near_core}.}
We will fix a constant $\d>0$ that is sufficiently small for various bounds to hold. We also take $B$  sufficiently large for various bounds to hold, including Lemma~\ref{pt} for $i\geq B$. Let $X_a = X_a(B)$ be the number of deletion trees $T$ in $D_B$ with $a$ vertices. Our goal is to show that there exists some constant $Q>0$ such that \whp\ $X_a=0$ for $a>Q\log n$, so in the following we may allow ourselves to assume that $a$ is greater than some sufficiently large constant.

To prove Lemma~\ref{near_core} we first observe that, by Proposition~\ref{lT}, we can assume $H_B$ is uniformly random conditional on its degree sequence. Since Lemma~\ref{near_core} asserts a  property to hold with high probability, it suffices to establish this property in the configuration model for $H_B$ (by Corollary~\ref{ccon}(a)). Moreover, recall that by Proposition~\ref{lT}(b), as $B$ is increased \whp\ the degree sequence of $H_B$ tends to that of the $r$-core.

Let $v_1,\ldots,v_a$ be the vertices of $T$. We first specify $d_i = \deg_T(v_i)$ for each $i$, noting that these degrees must sum to $2a-2$.  The number of ways to arrange these $a$ vertices into a tree with a specified degree sequence is $(a-2)!/\prod(d_i-1)!$
and there are $a$ choices for the root, $u$, of the tree. So, the number of choices for this step is:
\[\frac{a(a-2)!}{\prod(d_i-1)!} \enspace .\]

Next we choose the vertices of $T$.  Then for each edge of $T$, we choose a vertex-copy of each of its endpoints. To do so, for each $v_i$, we choose a copy of $v_i$ for each of the $d_i$ edges in $T$ incident with $v_i$.  If $\deg_{H_B}(v_i)=j$, then there are $j!/(j-d_i)!$ choices for the $d_i$ copies of $v_i$.  Since $\deg_{H_B}(v_i)\geq d_i$, the number of choices corresponding to $v_i$ is at most $\sum_{w: \deg_{H_B}(w)\geq d_i}\deg_{H_B}(w)!/(\deg_{{H_B}}(w)-d_i)!$. By Lemma \ref{ltail2}, this number is at most $(Y(d_i)+\hf\d)n$ where
\[
Y(d) =Y_B(d) = \sum_{j\geq d}\frac{j!}{(j-d)!} \r_j(B) \enspace .
\]

Furthermore,  if $d_i\leq\deg_{H_B}(v)\leq d_i+r-2$, then we can use $Y'(d_i)$ rather than $Y(d_i)$ where
\[
Y'(d)=Y'_B(d) =\sum_{j=d}^{d+r-2} \frac{j!}{(j-d)!}\r_j(B) \enspace .
\]
Using $Y'(d_i)$ instead of $Y(d_i)$ will be particularly useful when $d_i\leq 2$. By  Lemma \ref{pt}, for any $\d>0$ we can take $B=B(\d)>0$ sufficiently large, so that we must use $Y(d_i)$ for at most $\d a+3$ vertices $v_i$.  For convenience, we will assume $a>3/\d$ so we can take $\d a +3\leq 2\d a$. 

We will upper bound $\ex(X_a)$ by using  $Y(d_i)$  for every vertex $v_i$ with $d_i\geq 3$ and for exactly $2\d a$ vertices of degree $d\leq 2$. Let $t_1, t_2, t_3$ denote the number of vertices $v_i$ for which $d_i=1,d_i=2,d_i\geq 3$, respectively. We  note that  for sufficiently large $d$, $Y(d)$ is decreasing and so there is a constant $d^*$ such that for all $d\geq 3$, $Y(d)\leq Y(d^*)$. So, if we were to use $Y'(d)$ for every vertex of degree $d\leq 2$ then  the overall contribution of the $Y,Y'$ terms would be at most:
\[
[(Y'(1)+\hf\d)n]^{t_1} \cdot [(Y'(2)+\hf\d)n]^{t_2}  \cdot [(Y(d^*)+\hf\d)n]^{t_3} \enspace .
\]
We correct for the  $2\d a$ vertices of degree $d\leq 2$ for which we use $Y(d)$. To do so, we multiply  by the ${t_1+t_2 \choose 2\d a}\leq{a\choose 2\d a}$ choices for those vertices, and we multiply by $\Upsilon^{2\d a}$ where, for $\d$ sufficiently small,
\[
\Upsilon = \max\left(\frac{Y(1)+\hf\d}{Y'(1)+\hf\d},\frac{Y(2)+\hf\d}{Y'(2)+\hf\d}\right)=O(1)\enspace .
\]
 This brings the overall contribution of the $Y,Y'$ terms to at most: 
\[
{a\choose 2\d a} \Upsilon^{2\d a} [(Y(1)+\hf\d)n]^{t_1} [(Y(2)+\hf\d)n]^{t_2} [(Y(d^*)+\hf\d)n]^{t_3} \enspace . 
\]

Having chosen $d_1,\ldots,d_a$ and the vertices $v_1,\ldots,v_a$, we divide by the number of rearrangements
of those vertices; i.e. we multiply by 
\[\frac{1}{a!} \enspace .\]

Finally, we multiply by the probability that each of the $a-1$ pairs of vertex-copies corresponding to edges of $T$, lands in a hyperedge of the configuration. By Proposition \ref{p2arc}, no two such pairs lie in  the same hyperedge of $H_B$.
So, we can apply Lemma \ref{lconfig} to the $a-1$ specified pairs of vertex-copies and multiply by \[\left(\frac{k-1}{kE}\right)^{a-1}{e^{ka^2/(E-a)}}\]
 to get an overall bound, where $E$ is the number of edges in $H_B$.

Recall that for $c > c^*_{k,r}$, $\mu=\mu(c)$ denotes the larger of the two solutions of $f(\l) = c$. {By Proposition \ref{lT} and Lemma \ref{ltail}} for any $\d>0$,  we can take $B$ sufficiently large so that
\[
\left|kE- \mu\sum_{j\geq r-1}\frac{e^{-\mu} \mu^j}{j!}n \right|\leq \d n \enspace .
\]

Our key Lemma \ref{lc*} now yields that by taking $B$ sufficiently large, we can have $\d$ sufficiently small in terms of $\z$ that various bounds below hold, including
\begin{equation}\label{e3}
\left(\frac{e^{-\mu}\mu^r}{(r-2)!}+\d\right)\frac{k-1}{kE/n}<1-\frac{\z}{2} \enspace .
\end{equation}

By Lemma \ref{lT}, for any $\d>0$, we can take $B$ sufficiently large so that $Y'(1)\leq \d/2$ and $Y'(2)\leq \frac{e^{-\mu}\mu^r}{(r-2)!}+\d/2$. So, $Y'(1)+\hf\d,Y'(2)+\hf\d$ are bounded above by $\d$ and $\frac{e^{-\mu}\mu^r}{(r-2)!}+\d$, respectively. We let 
\[\Psi=2Y(d^*)>Y(d^*)+\hf\d \enspace ,\]
 for $\d$ sufficiently small.

Putting all this together, and recalling that $t_1+t_2+t_3=a$, yields
\begin{eqnarray}
\nonumber E(X_a) &\leq & 
{a\choose2\d a}
\Upsilon^{2\d a} \left(\frac{k-1}{kE}\right)^{a-1}
e^{{ka^2/(E-a)}} \\
&&\qquad\qquad\times
 \sum_{d_1+\cdots+d_a=2a-2}(\d n)^{t_1}
  \left[\left(\frac{e^{-\mu}\mu^r}{(r-2)!}+\d\right)n\right]^{t_2}  (\Psi n)^{t_3}
\frac{a (a-2)!}{a!\prod_{i=1}^a(d_i-1)!}\label{exa} \\
\nonumber& \le & {O(n/a)}
 e^{ka^2/(E-a)}
\left(\frac{\Upsilon^{2\d }}{(2\d)^{2\d}(1-2\d)^{1-2\d}}\right)^a
\left(\frac{k-1}{k{E/n}}\right)^{a}
\sum_{d_1+\cdots+d_a=2a-2}
{  \d ^{t_1} \left(\frac{e^{-\mu}\mu^r}{(r-2)!}+\d\right)^{t_2}  \Psi^{t_3} } \enspace .
\end{eqnarray}
Note that in the last line, we dropped the $\prod_{i=1}^a(d_i-1)!$ term.  We can afford to do so,
since this is equal to 1 for $d_i=1$ or 2, which are the most sensitive values.

 For $\d$ sufficiently small in terms of $\z$,
 \[
 \frac{\Upsilon^{2\d}}{(2\d)^{2\d}(1-2\d)^{1-2\d}}<1+\frac{\z}{10} \enspace .
 \]

Since {we are dealing with} the degree sequence of a tree, we have { $t_1>t_3$}. Since $\d<1$, we have $\sqrt{\d}^{t_1}<\sqrt{\d}^{t_3}$, yielding:
\begin{eqnarray*}
E(X_a)& < & O(n/a)e^{{ka^2/(E-a)}}
\left(1+\frac{\z}{10}\right)^a\\
&& \times\sum_{d_1+\cdots+d_a=2a-2}{\left(\sqrt{\d}\frac{k-1}{kE/n}\right)}^{t_1}
\left(\left[\frac{e^{-\mu}\mu^r}{(r-2)!}+{\d}\right]\frac{k-1}{k{E/n}}\right)^{t_2}
{\left(\sqrt{\d}\Psi\frac{k-1}{k{E/n}}\right)^{t_3} } \enspace .
\end{eqnarray*}
{Recalling that $E/n=\Omega(1)$ and $\Psi=O(1)$,} we choose $\d$  sufficiently small in terms of $\z$ so that
{\[\sqrt{\d}\frac{k-1}{kE/n}, \; \sqrt{\d}\Psi\frac{k-1}{kE/n}<\frac{\z}{100} \enspace .\]}
This and (\ref{e3}) yield

\begin{eqnarray*}
E(X_a)&\leq& O(n/a) e^{{ka^2/(E-a)}}\left(1+\frac{\z}{10}\right)^a\sum_{d_1+\cdots+d_a=2a-2}
\left(1-\frac{\z}{2}\right)^{t_2} \left(\frac{\z}{100}\right)^{a-t_2}.
\end{eqnarray*}

Now we fix $t_2$ and count the number of choices for $d_1,\ldots,d_a$.  There are ${a\choose t_2}$ choices for the values of $i$ with $d_i=2$. The remaining $a-t_2$ degrees sum to $2a-2-2t_2$.  The number of choices for {sequences of} $y$ non-negative integers that sum to $z$ is ${y+z-1\choose y-1}$, so the number of choices for these degrees is bounded by ${2(a-t_2)-3\choose a-t_2-1}<2^{2(a-t_2)-3}<4^{a-t_2}$. Thus,
\begin{eqnarray}
\nonumber E(X_a)&\leq&O(n/a)e^{{ka^2/(E-a)}}\left(1+\frac{\z}{10}\right)^a
\sum_{t_2=0}^a{a\choose t_2}4^{a-t_2}
\left(1-\frac{\z}{2}\right)^{t_2}\left(\frac{\z}{100}\right)^{a-t_2}\\
\nonumber&=&O(n/a)e^{{ka^2/(E-a)}}
\left(1+\frac{\z}{10}\right)^a\left(1-\frac{\z}{2}+\frac{\z}{25}\right)^a\\
\nonumber&<&O(n/a)e^{{ka^2/(E-a)}}\left(1-\frac{\z}{4}\right)^a\\
&<&O(n/a)\left(1-\frac{\z}{16}\right)^a \enspace ,\label{epb}
\end{eqnarray}
where the last inequality holds for all $a$ small enough that
$e^{{ka/(E-a)}}<1+\frac{\z}{4}$.  Thus, there are constants $Q,\xi>0$ such that $\ex (\sum_{a=Q\log n}^{\xi n} X_a)=o(1)$ and, therefore, \whp\ there are no deletion trees of size between $Q\log n$ and $\xi n$. Note now that $Q,\xi$ depend only on $\z,c,k,r$ and $\z$ depends only on $c,k,r$. 

Using Proposition \ref{lT}(a), we chose $B$ large enough that \whp\ $H_B$ contains fewer than $\xi n$ vertices outside of the $r$-core. Since  a deletion tree can have at most one vertex in the $r$-core, this implies that there are no deletion trees of size at least $\xi n$.   Therefore, \whp\ there are no deletion trees
in $H_B$ of size greater than $Q\log n$. Therefore, \whp\ for all $u\in D_B$, $|R^+(u)\cap H_B|\leq Q\log n$.
\proofend

\subsection{{Summing over a core flippable cycle for $r = 2$}}
Recall that for {Theorem~\ref{main_lemma}(b), we have} $r=2$; i.e., we consider $2$-cores for random $k$-uniform hypergraphs where $k \ge 3$.

Consider any {core} flippable cycle $C$ with vertices $u_1,\ldots,u_{\ell}$.  In our directed graph $D$, add edges from $u_j$ to $u_{j+1}$ for each $j$ (addition mod $\ell$).  Thus,
$R^+(u_1) = \cup_{j=1}^{\ell} R^+(u_j)$.  We modify the arguments from the proof of part (a) for this setting.

We define $T$ as in the previous section, this time rooted at $u_1$.  

We follow the proof of Lemma~\ref{pt}. Since $u_1,\ldots,u_{\ell}$ are the only 2-core vertices in $S$, we still have  $|S|\leq \theta n$. Since each $u_i$ has degree $2$ in the 2-core, it is easy to see that $\deg_{H_B}(u_i)=\deg_S(u_i)+1$. The proof of Lemma~\ref{pt} still holds, yielding
\[\deg_{H_B}(v)\leq\deg_T(v)+1 \enspace, \mbox{ for all but at most $\d|T|+3$ vertices }v\in T.\]
(In fact, this time we actually get $\d|T|+2$, but that is inconsequential.)

As in {Section~\ref{s5.1a}}, we bound the expected number of such trees of size $a$; $u_1$ is the root and hence plays the role of $u$ from {Section~\ref{s5.1a}}. This time, $T$ has the additional property that there is an  edge in $D$ from a vertex of $T$ (i.e. $u_{\ell}$) to $u_1$. To account for this additional property, we adjust (\ref{exa}) as follows: (i) multiply by the number of choices of one of the $a-1$ other vertices to be $u_{\ell}$;
(ii) choose vertex-copies for the extra edge from $u_{\ell}$ to $u_1$; (iii) adjust the term $\left(\frac{k-1}{kE}\right)^{a-1}{e^{ka^2/(E-a)}}$
which, by Lemma \ref{lconfig}, bounded the probability that the $a-1$ pairs of vertex-copies corresponding to edges of $T$ each landed in a hyperedge of the configuration.

For (ii), we use $Y(d(u_j)+1)$ instead of $Y(d(u_j))$ or $Y'(d(u_j))$ for $j=1,\ell$. For $j=1,\ell$, the adjustment for $u_{j}$ is an increase of a multiplicative factor of at most $(Y(d(u_{j})+1)+\hf\d)/(Y'(\deg(u_{j}))+\hf\d)<(Y(d^*)+\hf{\d})/(Y'(1)+\hf{\d})=O(1)$. So the overall effect
for (ii) is a multiplicative $O(1)$.

For (iii), the hyperedge containing $u_1,u_{\ell}$ is in the 2-core and so is distinct from the other $a-1$ hyperedges.  This results in another multiplicative factor of $\frac{k-1}{kE}$ to account for that edge,
when applying Lemma \ref{lconfig}.

The net result is to multipy $\ex(X_a)$ by $O(a/n)$, and so the bound on $\ex(X_a)$ in (\ref{epb}) becomes $O(1)\left(1-\frac{\z}{16}\right)^a$. Summing over all $a$ yields that the expected number of {core} flippable cycles $C$ such that $|\bigcup_{u\in C}R^+(u)\cap H_B|>{\xi}(n)$ is $o(1)$ for any ${\xi}(n)\rightarrow\infty$, {in particular for $\xi(n)=O(\log n)$}.
Proposition \ref{pns} and Lemma \ref{lst} {yield Theorem~\ref{main_lemma}(b)}.
\proofend

\section{Proof of Theorem~\ref{main_lemma} below the $r$-core threshold}\label{sec:below}
Recall that Theorem~\ref{main_lemma} is already known for $r=k=2$, so we will assume $r+k>4$.
As in the case for $c>c^*_{k,r}$, we will carry out a large but fixed number, $I$, of rounds of the parallel $r$-stripping process, ending up with a hypergraph $H_I$. Because we are below the  $r$-core threshold, this will delete all but  a very small, albeit linear, number of vertices.  Proposition \ref{pdeg} asserts that the remaining hypergraph is uniformly random conditional on its degree sequence. We will determine this degree sequence and apply the technique from  \cite{mr} to show that the maximum component size in the remaining hypergraph has size $O(\log n)$. Thus, for every $v$, we must have $|R^+(v)\cap H_I|=O(\log n)$. Proposition \ref{pns} and Lemma \ref{lst} then imply that $|R^+(v)|=O(\log n)$ as required.

Let $\mathrm{Po}(\mu)$ denote a Poisson variable with mean $\mu$. Recursively define the following quantities:
\begin{eqnarray*}
\phi_0 & = & 1\\
\l_t & = & c\phi_t^{k-1}/(k-1)! \\
\phi_{t} & = & \Pr[\mathrm{Po}(\l_{t-1})\geq r-1] \enspace .
\end{eqnarray*}

Write  $P(\mu,j)=\Pr[\mathrm{Po}(\mu) = j]$.

\begin{lemma}\label{lem:core_deg} For any constants $d,t$, the number of vertices of degree $d$ after $t$ rounds of the parallel $r$-stripping process, \whp\ is $\r_t(d)n+o(n)$ , where 
\begin{equation*}
\r_t(d) =
\begin{cases} P(\l_t,d) & \text{for  $d \ge r$,}
\\
\\
P(\l_t,d)\cdot
\Pr\left[
\mathrm{Po}(\l_{t-1}-\l_t)
\geq r-d
\right]
&\text{for  $d < r$.}
\end{cases}
\end{equation*}
\end{lemma}
\begin{proof} We consider a branching process  introduced in \cite{psw} and analyze it as in \cite{mmcore}.
Consider any hypergraph $H$ and any vertex $v\in H$. For each $0\leq i\leq t+1$, let $L_i(v)$ be the  vertices of distance at most $i$ from $v$ (thus $L_0(v)=\{v\}$). For any $u\in L_i(v)$ with $0\leq i\leq t$,  a {\em child edge} of $u$ is an edge containing $u$ and $k-1$ members of $L_{i+1}(v)$; thus if the distance $t+1$ neighbourhood of $v$ induces a hypertree, then all but at most one of the edges containing $u$ are child edges of $u$. 

We define the process STRIP$(v,t)$ as follows: \smallskip

For $j$ from $t$ down to $1$ do\\
\indent \indent Remove all vertices in $L_j(v)$ with fewer than $r-1$ child edges; \\
\indent \indent Remove all edges that contain a removed vertex. \medskip

\noindent Let $X_t$ denote the number of child edges of $v$ that survive STRIP$(v,t)$, and let $Y_t$ denote 
the number of child edges of $v$ that survive STRIP$(v,t-1)$ but not STRIP$(v,t)$.
If the hypergraph induced by the vertices in $L_{t+1}(v)$ induces a hypertree, then we see that
\begin{itemize}
\item[(A)]
For $d\geq r$: $v$ survives {the first} $t$ rounds of the parallel $r$-stripping process, and has degree $d$ in what remains iff $X_t=d$.
\item[(B)]
For $1\leq d < r$: $v$ survives {the first}  $t$ rounds of the parallel $r$-stripping process, and has degree $d$ in what remains iff $X_t=d$ and $Y_t\geq r-d$.
\end{itemize}

To analyze STRIP$(v,t)$ on $H=H_k(n,p=c/n^{k-1})$, we make use of the fact that \whp\ the distance $t+1$ neighbourhood of $v$ induces a hypertree, and so both (A) and (B) hold.

We will argue by induction on $t$ that the probability a particular child $u$ of $v$ survives STRIP$(v,t)$ is $\phi_t+o(1)$.  Suppose $u\in L_1(v)$ and $w\in L_2(v)$ is in a child edge of $u$.
Note that $w$ survives STRIP$(v,t)$ iff $w$ survives STRIP$(u,t-1)$ the probability of which, by induction on $t$,
is easily seen to be $\phi_{t-1}+o(1)$.  It follows that the expected number of child edges of $u$ that survive STRIP$(v,t)$ is  $\frac{c}{n^{k-1}}{n-O(1)\choose k-1}(\phi_{t-1}+o(1))^{k-1}=\l_{t-1}+o(1)$.   Standard arguments show that for any fixed $i$ the probability that the number of such edges is $i$ is $P(\l_{t-1},i)+o(1)$ (we elaborate more below on similar arguments for $X_t,Y_t$).  Therefore, the probability that $u$ survives
STRIP$(v,t)$ is $\phi_t+o(1)$, thus completing the induction.
By the same argument, $\ex(X_t)=\l_t+o(1)$. Noting that $Y_t=X_{t-1}-X_t$, this yields $\ex(Y_t)=\l_{t-1}-\l_t+o(1)$. 

Consider any child edge $e$ of $v$ in $L_{t+1}(v)$. Whether $e$ counts towards $X_t,Y_t$ or neither is determined entirely by the subtrees of $L_{t+1}(v)$ rooted at the vertices of $e$ other than $v$. In other words, $X_t,Y_t$ are determined by the edges containing $v$ in $H_k(n,p=c/n^{k-1})$, and some local information about each edge where the information for any two edges is \whp\ disjoint. Also, no edge counts towards both $X_t$ and $Y_t$.  From this, it is straightforward to show, using e.g., the {Method of Moments,} (see \cite{jlr}) that the joint distribution of $X_t,Y_t$ is asymptotic to independent Poisson variables; specifically, for any fixed integers $x,y$, $\pr(X_t=x \wedge Y_t=y)$ is $o(1)$ plus the probability that two independent Poisson variables with means $\ex(X_t),\ex(Y_t)$ are equal to $x,y$.

(A) and (B) now yield that the probability that $v$ survives the first $t$ rounds of the parallel stripping process and has degree $d$ in $H_t$ is $\r_t(d)+o(1)$, and
so the expected number of such vertices is $\r_t(d)n+o(n)$.  The lemma now follows as in \cite{mmcore} from a straightforward concentration
argument, e.g., a second moment calculation.  We omit the details.
\end{proof}

The main  result of \cite{mr} {states: Consider a random graph on a fixed degree sequence where $\Lambda(d) \cdot n + o(n)$ vertices have degree $d$, and  where the degree sequence satisfies certain {\em well-behaved} conditions. If}
\begin{equation}\label{molreeds}
\sum_{d\geq1}  d(2-d) \Lambda(d)> 0 \enspace ,
\end{equation}
and
then \whp\ all connected components have size $O(\log n)$. A simple adaptation of the proof in~\cite{mr} provides a generalization to hypergraphs. Specifically, for $k>2$ it suffices to replace $d(2-d)$ in~\eqref{molreeds}  with 
$$
f_k(d) = d  [1-(d-1) (k-1)] \enspace .
$$
{Proposition~\ref{pdeg} allows us to model $H_t$ as a random hypergraph on  degree sequence $\r_0(t),\r_1(t),...$. Using Lemma \ref{ltail2}, it is straightforward to verify that this degree sequence satisfies the well-behaved conditions from \cite{mr}, and so deduce that if}
\begin{equation}\label{eq:MRc}
\sum_{d\geq1} \r_t(d) f_k(d) > 0  \enspace ,
\end{equation}
then \whp\ all  components of $H_t$ have size $O(\log n)$.

Since $\Pr[\mathrm{Po}(\l) \ge r-1]$ is a strictly increasing function of $\l$, the sequences $\phi_t,\l_t$ are strictly decreasing. If they do not tend to zero, then there must be a positive fixed point to the recursion defining them, i.e., a positive solution to
\[\l  =  c\Pr[\mathrm{Po}(\l)\geq r-1]^{k-1}/(k-1)! \enspace .\]
Rearranging yields
\[c=(k-1)!\l/\Pr[\mathrm{Po}(\l)\geq r-1]^{k-1} \enspace.\]
Recall now that $c_{k,r}^*$ was defined in~\eqref{krthreshold} as the smallest value of $c$ for which there is such a solution. Since $c<c_{k,r}^*$, we can conclude that $\phi_t,\l_t \to 0$ as $t \to \infty$ and we can develop the following asymptotics {in $t$}, {using $O_t()$ and $\Theta_t()$ to denote asymptotics are with respect to $t$:}
\begin{equation}\label{eq:lasymp}
\l_t = \frac{c}{(k-1)!}\, \phi_t^{k-1}  = \frac{c}{(k-1)!} \, \left(\Pr[\mathrm{Po}(\l_{t-1}) \ge r-1]\right)^{k-1} = \Theta_t(\l_{t-1}^{(k-1)(r-1)}) \enspace .
\end{equation}

{Let $\l :=\l_t$ and $\theta :=\l_{t-1}-\l_t$. Since  $(k-1)(r-1) \ge 2$ for $k+r>4$, we see that \eqref{eq:lasymp} implies $\l = o_{t}(\theta)$.}

We apply Lemma~\ref{lem:core_deg}, noting that as $\theta \to 0$,  $\Pr[\mathrm{Po}(\theta) \ge r-d] \to P(\theta,r-d)$. Therefore, as $t \to \infty$, inequality~\eqref{eq:MRc} is equivalent to
\begin{equation}\label{masteropoulos}
(1+o_{t}(1))\sum_{d = 1}^{r-1}
P(\l,d) P(\theta,r-d)
f_k(d)
+\sum_{d = r}^{\infty}
P(\l,d)
f_k(d)
> 0
\enspace .
\end{equation}
Note that $f_k(1)=1$ and $f_k(d)\leq 0$ for $d\geq 2$. So the first sum in (\ref{masteropoulos}) is {at least}
\begin{equation}\label{gagaga} 
P(\l,1) P(\theta,r-1)-\sum_{d = 2}^{r-1} P(\l,d) P(\theta,r-d)|f_k(d)| \enspace .
\end{equation}
For $1\leq d\leq r-1$, we have $f_k(d)=O_t(1)$, so the first term {in~\eqref{gagaga}} is
$\Theta_t(\l \theta^{r-1})$ {while} the sum in~\eqref{gagaga} is $\Theta_t(\sum_{d = 2}^{r-1}\l^d\theta^{r-d})=\Theta_{t}(\l^2 \theta^{r-2})$,
since $\l = o_{t}(\theta)$.  Therefore, the first sum in (\ref{masteropoulos}) is positive
and of order $\Theta_t(\l \theta^{r-1})$.

At the same time, {since $-f_k(d) = k d^2 -d^2-kd < k d^2$ we get}
\[
- \sum_{d = r}^{\infty} P(\l,d)  f_k(d) \le k
\sum_{d = r}^{\infty} P(\l,d)  {d^2}  = O_{t}(\l^r) \quad , \text{ as $\l\to 0$}\enspace .
\]

{Thus, the first sum in~\eqref{masteropoulos} is positive $\Theta_{t}(\l \theta^{r-1})$, whereas the second sum is $O_{t}(\lambda^r)$. Since $\l = o_{t}(\theta)$ it follows that (\ref{eq:MRc}) holds for $t$ sufficiently large and, therefore, for $I$ sufficiently large, every component of $H_I$ has size $O(\log n)$.} Theorem~\ref{main_lemma}(b) now follows
from Proposition \ref{pns} and Lemma \ref{lst}.
\proofend

\section{Proof of Theorem \ref{t2}}\label{spt2}
Given a solution, {recall} that a  set $S$ of variables is {\em flippable} if changing the assignment of every variable in $S$ results in another solution.  Note that flippable sets can be characterized in terms of the underlying hypergraph.

\begin{proposition}\label{pflip}
$S$ is flippable iff every hyperedge contains an even number of members of $S$.
\end{proposition}

So we define:
\begin{definition} A {\em flippable set} in a hypergraph, $H$, is a {nonempty} set of vertices, $S$, such that every edge in $H$ contains
 an even number of vertices of $S$.
\end{definition}

Recalling Definition \ref{flippable_def}, we see that a flippable cycle is a flippable set. A flippable set is {\em minimal} if it does not contain a flippable proper subset. { Note that every flippable set contains a minimal flippable subset.}

\begin{lemma}\label{lt2}
Let $H$ be a random $k$-uniform hypergraph $H_k(n,p)$, where $p =c/n^{k-1}$. For every $c>c^*_{k,2}$ there exists $\a >0$ such that \whp\ every minimal flippable set in the {hypergraph induced by the} 2-core of $H$ either is a {core} flippable cycle or has size at least $\a n$.
\end{lemma}

Lemma \ref{lt2} follows immediately from Lemma \ref{lex} below, and yields Theorem \ref{t2} as follows:

\begin{proof}[Proof of Theorem \ref{t2}]
Consider any two solutions $\s_1,\s_2$ in different clusters.  Let $S$ be the variables in the 2-core on which these solutions disagree.
Thus, $S$ is a flippable set in the hypergraph induced by the 2-core. Remove all core flippable cycles from $S$, and let $S'$ be what remains {(recall from Definition~\ref{flippable_def} that a flippable cycle is a set of vertices).} Lemma~\ref{flip_cycles}(b) implies that \whp\ every 2-core hyperedge contains an even number of vertices that lie in core flippable cycles. Therefore $S'$ must also be a flippable set in the hypergraph induced by the 2-core. By the definition of clusters, $S'\neq\emptyset$ as otherwise $\s_1,\s_2$ would be cycle-equivalent.
Let $S''$ be a mimimal flippable subset of $S'$. Since $S''$ contains no {core} flippable cycles, Lemma \ref{lt2} implies that \whp\ $|S''|\geq \a n$.
Therefore $|S|\geq \a n$ and so $\s_1,\s_2$ differ on at least $\a n$ variables.

Any sequence $\s,\s',\ldots,\t$ where $\s,\t$ are in different clusters must contain two consecutive solutions that are in different clusters.
As argued above, those two solutions differ on at least $\a n$ variables.  It follows that if $\s,\t$ are in different clusters then
$\s,t$ are not $\a n$-connected.
\end{proof}

If we could show (deterministically) that the hypergraph induced by any minimal flippable set in a 2-core that is not a {core} flippable cycle is sufficiently dense, then Lemma \ref{lt2} would follow by a rather standard argument.  Unfortunately, there is no useful lower bound on the density, mainly because of the possibility of very long 2-linked paths in $S$ (defined below).  Instead, we follow an approach akin to that of~\cite{ms}, forming a graph $\G(S)$ by contracting those long paths, and making use of the fact that $\G(S)$ is dense (Lemma \ref{ldense}). The main difference from~\cite{ms} is that here we need to work in the configuration model.

To prove Lemma \ref{lt2}, we first require a few definitions.  Note that these concern any hypergraph, not just a {2-core of a random hypergraph}.  

A hyperedge is \emph{simple} if it is not a loop, i.e., if it does not contain any vertex more than once.

\begin{definition} \label{d2l}
Let $\calh$ be  a $k$-uniform hypergraph. A {\em 2-linked path} {$P$} of a set $S\subseteq V(\calh)$ is a set  of vertices $v_0,\ldots,v_t \in S$ and simple hyperedges $e_1,\ldots,e_{t}$, {where $t\geq 1$}, such that
\begin{itemize}
\item[(i)] $v_0,\ldots,v_t$ are all distinct except that {when $t\geq 2$ we allow} $v_0=v_t$. {(Note that if $v_0=v_t$ then these vertices actually form a cycle and so {\em 2-linked path} is somewhat of a misnomer.)}
\item[(ii)] Each $e_i$ contains $v_{i-1},v_{i}$ and no other vertices  of $S$.
\item[(iii)] $v_1,\ldots,v_{t-1}$ all have degree 2 in $\calh$; {i.e.\ they do not lie in any edges outside of $P$.}
\item[(iv)]
{If $v_0=v_t$ then  $\deg_{\calh}(v_0)>2$.
If $v_0\neq v_t$ then} {$S$ is maximal w.r.t.\ (ii) and (iii); i.e.\  each} of $v_0,v_t$ either has degree $\neq 2$ in $\calh$, or lies in a hyperedge $e{\notin P}$ with $|e\cap S|\neq 2$.
\end{itemize}
We call $v_0,v_t$ the \emph{endpoints} of the path and $v_1,\ldots,v_{t-1}$ its \emph{connecting vertices}.
\end{definition}

Note that if $v_0=v_t$ then by (iv), $\deg_{\calh}(v_0)>2$ and hence $v_0,\ldots,v_t$ do not form a flippable cycle.

\begin{figure}[h]\label{fig:2-linked}
\begin{center}
    \caption{2-linked paths with $t=3$.  On the left $v_0\neq v_3$, while on the right $v_0=v_3$. Vertices in $S$ are marked with a square.}
\end{center}
\end{figure}

\begin{definition}\label{d38} We say that $S\subseteq V({\calh})$ is a {\em linked set} if (i) $S$ does not contain a flippable cycle as a subset, (ii) no hyperedge of ${\calh}$ contains exactly one element of $S$ and (iii)  every hyperedge $e$ of ${\calh}$ with $|e\cap S|=2$ is in a 2-linked path of $S$.
\end{definition}

\begin{proposition}\label{p2link}  Suppose $S$ is a flippable set {in a hypergraph where all hyperedges are simple, and $S$} does not contain a flippable cycle as a subset. Then $S$ is a linked set.
\end{proposition}

\begin{proof}
By Proposition \ref{pflip}, we only need to check condition (iii) of Definition~\ref{d38}.  Consider any hyperedge $e$ with $|e\cap S|=2$. {Since $e$ is simple,} either $e$ itself forms a 2-linked path in $S$, or it is easily seen that $e$ can be extended into such a path, unless $e$ lies in a flippable cycle.
\end{proof}

{\bf Remark:} It is easy to see that in any Uniquely Extendible CSP, the set of disagreeing variables of any two solutions must be a flippable set. Since Proposition \ref{p2link} was derived by only considering the underlying hypergraph (and not the specific constraints), it applies to any UE CSP. Therefore, our Theorem~\ref{t2} extends readily to every UE CSP since its proof amounts to proving that for some constant $\a>0$, all linked sets are either flippable cycles or contain at least $\a n$ variables.\medskip

Given a linked set, $S$, we consider the mixed hypergraph {(containing both hyperedges and normal edges)} $\G(S)$ formed as follows:
\begin{enumerate}
\item[(a)] The vertices of $\G(S)$ are the endpoints of the 2-linked paths in $S$ along with all vertices of $S$ that do not lie in any 2-linked paths.
\item[(b)] There is an edge in $\G(S)$ between the endpoints of each 2-linked path in $S$. That edge is a loop if the
two endpoints are the same vertex, and so $\G(S)$ is not necessarily simple.
\item[(c)] For every hyperedge $e$ of ${\calh}$ with $|e\cap S|>2$, $e\cap S$ is a hyperedge of $\G(S)$.
\end{enumerate}

 Thus $V(\G(S))\subseteq S$, and since no hyperedge of $C$  contains exactly one element of $S$, for every $v\in V(\G(S))$ we have $\deg_{\G(S)}(v)=\deg_{\calh}(v)$. Any vertex of $S$ that is not in $\G(S)$ is a connecting vertex of a 2-linked path in $S$.

\begin{proposition}\label{p16} If $S$ is a non-empty linked set, then $1\leq|\G(S)|\leq |S|$.
\end{proposition}

\begin{proof}
Any vertex of $S$ that is not in $\G(S)$ is a connecting vertex of a 2-linked path in $S$. The endpoints of that 2-linked path are in $\G(S)$. Thus $|\G(S)|\geq1$. The rest follows from the fact that every vertex of $\G(S)$ is a vertex of $S$.
\end{proof}

Note that $\G(S)$ contains hyperedges of size between 2 and $k$.
For each $2\leq i\leq k$, we define $\ell_i$ to be the number of $i$-edges in $\G(S)$.

\begin{lemma}\label{ldense} If every vertex in $\calh$ has degree at least 2 then
$\sum_{i=2}^k (i-1)\ell_i\geq (1+\frac{1}{2k}) |V(\G(S))|$.
\end{lemma}

\begin{proof}
As we said above, every $v\in V(\G(S))$ has the same degree in $\G(S)$ as it does in $\calh$. Thus $\G(S)$ has minimum degree at least 2.  Consider any $v$ of degree 2 in $\G(S)$.  Then $v$ has degree 2 in $\calh$ and hence cannot be the endpoint of a 2-linked path in $S$, unless $v$ lies in at least one hyperedge of $\calh$ containing more than 2 members of $S$.  It follows that $v$ lies in at least one hyperedge of $\G(S)$ of size greater than 2. Therefore, at most $\sum_{i=3}^ki\ell_i<k\sum_{i=3}^k\ell_i$ vertices of $\G(S)$ have degree 2, and so letting $Z$ denote the number of vertices with degree at least 3 in $\G(S)$, we have
\[|V(\G(S))|\leq Z+k\sum_{i=3}^k\ell_i\leq k\left(Z+\sum_{i=3}^k\ell_i\right).\]

By the handshaking lemma, $\sum_{i=2}^k i\ell_i =\sum_v \deg_{\G(S)}(v)$. Therefore,

\begin{eqnarray*}
\sum_{i=2}^k (i-1)\ell_i & = & {\hf\sum_v \deg_{\G(S)}(v)+{\sum_{i=2}^k (i/2-1) \ell_i }} \\
& \ge &\hf\sum_v \deg_{\G(S)}(v) +
\hf\sum_{i=3}^k\ell_i \\
&=& \sum_v 1 + \sum_v \hf(\deg_{\G(S)}(v)-2)    +\hf\sum_{i=3}^k\ell_i \\
&\geq& |V(\G(S))|+\hf Z +\hf\sum_{i=3}^k\ell_i \quad , \quad\mbox{since $\deg_{\G(S)(v)}\geq 2$ for all $v$}  \\
&\geq&\left(1+\frac{1}{2k}\right)|V(\G(S))| \enspace .
\end{eqnarray*}
\end{proof}

Let $C$ be the 2-core of $H=H_k(n,p)$. We will apply Lemma \ref{ldense} with $\calh=C$ to prove:

\begin{lemma}\label{lex} There exists $\a>0$ such that \whp\ $C$ has no {non-empty} linked set of size less than $\a n$.
\end{lemma}

Lemma \ref{lt2} follows immediately from Lemma \ref{lex} and Proposition~\ref{p2link} {(since  $H_k(n,p)$ contains only simple hyperedges).}
The proof of Lemma \ref{lex} will be reminiscent of the proof of Lemma \ref{l22}, but significantly more complicated because (i) we are working in the configuration model and (ii) where we had $\ell_2$ 2-edges in Lemma \ref{lex}, we have $\ell_2$  2-linked paths here. First, we provide a technical lemma.

\begin{lemma} \label{lc} For any integers $a,t$, given a set of $a$ vertices in $H=H_k(n,p)$, {with $p=c/n^{k-1}$}
the probability that their total degree exceeds $tkca$ is at most $\left(e/t\right)^{act}$.
\end{lemma}

\begin{proof}
Given a set $A$ of $a$ vertices, let $E_A$ denote the number of hyperedges containing at least one member of $A$.  The total degree in $A$ is at most $kE_A$.  The number of potential edges in $E_A$ is at most
$a{n\choose k-1}<an^{k-1}$, and so $E_A$ is dominated from above by $\mathrm{Bin}(an^{k-1},c/n^{k-1})$ and using ${n \choose z} \le (ne/z)^z$ we get
\[
\pr\left[
\mathrm{Bin}(an^{k-1},c/n^{k-1}) > act
\right]
<{a n^{k-1}\choose act}\left(\frac{c}{n^{k-1}}\right)^{act} < (e/t)^{act}.\]
\end{proof}

{\em Proof of Lemma \ref{lex}.} By Corollary \ref{ccon}, we can work in the configuration model.
Let $\cald$ be the degree sequence of  $C$. Recalling Definition~\ref{mu_def}, Proposition~\ref{ldeg} and our key Lemma \ref{lc*}, we have \whp\
\begin{enumerate}
\item[(i)] $\cald$ has total degree $\g n+o(n)$, where $\gamma = \mu \Psi_r(\mu)$,
\item[(ii)] $\cald$ has $\la_2 n+o(n)$ vertices of degree 2,  where $\lambda_2 = e^{-\mu} \mu^2/2$,
\item[(iii)]  there exists $\z >0$ such that
$2(k-1)\la_2<(1-\z)\g$.
\end{enumerate}

For each $a\geq 1$,  let $X_{a}$ denote the number of linked sets $S$ in $C$ for which  $|\G(S)|=a$ and let $X=\sum_{a= 1}^{\a n} X_a$. Define
\[\call_a=\left\{(\ell_2,\ldots,\ell_k) : \left(1+\frac{1}{2k}\right) a\leq \sum_{i=2}^k (i-1)\ell_i\leq \left(1+\frac{1}{2k}\right) a+(k-1)\right\}.\]

By Lemma \ref{ldense}, for any linked set $S$ in $C$ with $|\G(S)|=a$, there is some $(\ell_2,\ldots,\ell_k)\in\call_a$
so that $\G(S)$ contains {\em at least} $\ell_i$ $i$-edges for each $i$.

{
To bound $E(X_a)$, we begin by choosing $a$ vertices, $A\subseteq V(C)$ and sum over all $t\geq 0$ of the probability that  their total degree in $C$ lies in the range $(tkca,(t+1)kca]$. For each $t$, we upper bound this last probability by the probability that their total degree in $H$ lies in $(tkca,\infty]$. Moreover, to sum over all subsets $A \subseteq V(C)$ we overcount by summing instead over all  $A\subseteq V(H)$, and using Lemma \ref{lc}. Of course, if such a set is not a subset of $C$ then the probability of it contributing to $X_a$ is zero, and so this provides an upperbound on $\ex(X_a)$. This yields:
\[{n\choose a}\sum_{t\geq 0}{(\frac{e}{t})^{tca}}.\]

Given $A$, we sum over all possibilities for the values of $(\ell_2,\ldots,\ell_k)\in\call_a$. }   For each $2\leq i\leq k$, we choose $\ell_i$ $i$-sets of vertex-copies belonging to vertices of $A$. If the total degree of $A$ is in $(tkca,(t+1)kca]$ then the number of choices for these $\ell_i$ $i$-sets is at most
\[
\left(\frac{((t+1)kca)^i}{i!}\right)^{\ell_i}/\ell_i! <\frac{((t+1)kca)^{i\ell_{i}}}{\ell_i!} \enspace .
\]

Denote the $\ell_2$ {2-sets as $\{u_1,w_1\},\ldots,\{u_{\ell_2},w_{\ell_2}\}$.}  For each $i=1,\ldots,\ell_2$, we select $j_i\geq 0$, the number of connecting variables in the 2-linked path from $u_i$ to $w_i$ in $S$,  we choose {the} $j_i$ degree two { connecting variables for that path}, and we choose one of the two possible orientations of the vertex-copies of each of those connecting variables.  Let $J=j_1+\cdots+j_{\ell_2}$, be the number of connecting variables selected.  Let $L=\la_2n+o(n)$ be the number
of degree 2 vertices in $C$.  Then the total number of choices for the connecting vertices and the orientations of their copies is at most 
\[\prod_{i=1}^J 2(L-i+1).\]

Next, we apply Lemma \ref{lconfig} to bound the probability that the $\ell_3+\cdots+\ell_k$ sets of size at least 3 all land in
hyperedges of the configuration and that for each $i=1,\ldots,\ell_2$, the {\em first} pair in the 2-linked path, i.e., $u_i$ and the first copy of the first of the $j_i$ connecting variables, lands in a hyperedge of  the configuration. Note that $\ell_2+\cdots+\ell_k\leq \sum_{i=2}^k (i-1)\ell_i <2a+k-1<2a+o(n)$, by the definition of $\call_a$.  By {assuming $a < \a n $ for some sufficiently small $\a$}, we get $\g n +o(n) -2a>\hf\g n$. Therefore, Lemma \ref{lconfig} yields that this probability is at most
\begin{eqnarray*}
& & \exp
\left(
\frac{k(\ell_2+\cdots\ell_k)^2}{\hf\g n }
\right)
\prod_{i=2}^k\left(\frac{(k-1)(k-2)\cdots(k-i+1)}{(\g n+o(n))^{i-1}}\right)^{\ell_i} \\
&<&{
\exp
\left(
\frac{8ka^2}{\g n }
\right)
\prod_{i=2}^k\left(\frac{k}{\g n}\right)^{(i-1)\ell_i}\enspace .}
\end{eqnarray*}

Following the analysis of Lemma \ref{lconfig}, we have now exposed $\ell_2+\cdots+\ell_k$ hyperedges of  the configuration. Let $\Lambda$ be the number of unmatched vertex-copies remaining. Since $\ell_2+\cdots+\ell_k<2a+k-1$, we have $\Lambda\geq \g n -2ka+o(n)$.  If the other vertex-copies required for the 2-linked paths are still unmatched, then we continue; else we halt observing that in this case, the set of choices made so far cannot lead to a linked set on the chosen vertices.

There are $J$ pairs of vertex copies that each need to be in a hyperedge of  the configuration  in order to complete the 2-linked paths.  Following the same argument as in Lemma \ref{flip_cycles}, the probability of this happening is at most
\[\prod_{i=1}^J\frac{k-1}{\Lambda-k(i-1)}.\]
Applying (iii) above,  and taking $a<\a n$ for $\a$ sufficiently small in terms of $\g,\la_2$, we obtain:
\[
\frac{2(k-1)L}{\Lambda}<\frac{2(k-1)\la_2 n +o(n)}{\g n -{2ka} +o(n)}<1-\frac{\z}{2} \enspace .
\]
Thus, since $2(k-1)L \le \Lambda$ (by the previous line) and  $k\leq 2(k-1)$, we have $\frac{2(k-1)(L-(i-1))}{\Lambda-k(i-1)}<1-\frac{\z}{2}$ for each $i$, leading to
\begin{eqnarray*}
\ex(X_{a})&<&{n\choose a}\sum_{t\geq 0}{(\frac{e}{t})^{tca}}\sum_{\ell_2,\ldots,\ell_k\in\call_a}\sum_{j_1,\ldots,j_{\ell_2}\geq 0}e^{8ka^2/(\g n)}
\left(\prod_{i=2}^k\frac{({{(t+1)}kca})^{i\ell_i}}{\ell_i!}\right)\\
&&\times \left(\prod_{i=2}^k{\left(\frac{k}{\g n}\right)^{(i-1)\ell_i}}\right)
\left(\prod_{i=1}^J\frac{2(k-1)(L-(i-1))}{\Lambda-k(i-1)} \right)\\
&<&\left(\frac{en}{a}\right)^a\sum_{t\geq 0}{(\frac{e}{t})^{tca}}
\sum_{\ell_2,\ldots,\ell_k\in\call_a}  e^{8ka^2/(\g n)}
\left(\prod_{i=2}^k\frac{({kca})^{\ell_i}}{\ell_i!}
\left(\frac{{k^2ca}}{\g n}\right)^{(i-1)\ell_i}{(t+1)^{i\ell_i}}\right)\\
&&\times
\sum_{ j_1,\ldots,j_{\ell_2}\geq0}(1-\z/2)^J\enspace .
\end{eqnarray*}
Since $J=j_1+\cdots+j_{\ell_2}$, we have $\sum_{ j_1,\ldots,j_{\ell_2}\geq0}(1-\z/2)^J=\left(\sum_{j \geq 0}(1-\z/2)^j\right)^{\ell_2}
=(2/\z)^{\ell_2}$, yielding:
\[\ex(X_a)<\left(\frac{en}{a}\right)^a e^{8{k}a^2/\g n}\sum_{\ell_2,\ldots,\ell_k\in\call_a}
\left(\frac{{k^2ca}}{\g n}\right)^{\sum_{i=2}^k(i-1)\ell_i}
\left(\prod_{i=2}^k\frac{(kca)^{\ell_i}}{\ell_i!}\right){ (\frac{2}{\z})^{\ell_2}}
\sum_{t\geq 0}{(\frac{e}{t})^{tca}}{(t+1)}^{\sum_{i=2}^ki\ell_i}.
\]

{By our choice of $\mathcal{L}_a$}
\[\ell_2\leq\sum_{i=2}^k(i-1)\ell_i\leq(1+\inv{2k})a+k-1,\]
\[\sum_{i=2}^k i\ell_i\leq2\sum_{i=2}^k(i-1)\ell_i\leq3a+2k.\]
Thus, we obtain $(2/\z)^{\ell_2}<Z_1^a$ for constant $Z_1=Z_1(c)$
and, since $(e/t)^{tc}(t+1)^{3+2k}$ is decreasing for large $t$, we have
\[\sum_{t\geq 0}{(e/t)^{tca}}{(t+1)}^{\sum_{i=2}^ki\ell_i}<{\sum_{t\geq 0}(e/t)^{tca}{(t+1)}^{3a+2k}
<\sum_{t\geq 0}\left((e/t)^{tc}{(t+1)}^{3+2k}\right)^a<}Z_2^a \enspace ,\]
for constant $Z_2=Z_2(c)$. Also using $a\leq n$ we obtain:
\begin{eqnarray*}
\ex(X_{a})&<&\left(\frac{en}{a}\right)^a e^{8{k}a/\g}({Z_1Z_2})^a\left(\frac{k^2ca}{\g n}\right)^{\left(1+\inv{2k}\right)a+k-1}
\sum_{\ell_2,\ldots,\ell_k\geq0}\prod_{i=2}^k\frac{(kca)^{\ell_i}}{\ell_i!} \\
&<&O(1)\left(e{Z_1Z_2}e^{8{k}/\g}\left(\frac{k^2c}{\g}\right)^{1+\inv{2k}}\right)^a\left(\frac{a}{n}\right)^{a/2k}
\left(\sum_{\ell\geq 0}\frac{(kca)^{\ell}}{\ell!}\right)^{k-1} \enspace .
\end{eqnarray*}
Applying $\left(\sum_{\ell\geq 0}\frac{(kca)^{\ell}}{\ell!}\right)^{k-1}=e^{kca(k-1)}$ we obtain:
\[\ex(X_a)<O(1)\left(e{Z_1Z_2}e^{8{k}/\g}(k^2c/ \g)^{1+\inv{2k}}e^{ck(k-1)}\right)^a\left(\frac{a}{n}\right)^{a/2k}
<Y^a\left(\frac{a}{n}\right)^{a/2k},
\]
for a constant $Y=Y(\g,\la_2,\z,b,\xi)$ that does not depend on $a$, so long as $a<\a n$ for  sufficiently small $\a>0$.  This yields $\ex(\sum_{a=1}^{\sqrt{n}} X_a)=o(1)$. Moreover, for all $\a$ sufficiently small, $\ex(X_a)<2^{-a}$. Therefore, $\ex(\sum_{a\geq\sqrt{n}} X_a)=o(1)$ and, thus, $\ex(X)=o(1)$.

Therefore, \whp\ there is no 2-linked set $S$ with $1\leq|\G(S)|\leq \a n$.  The lemma follows
from Proposition~\ref{p16}.
\proofend


\begin{thebibliography}{99}

\bibitem{ABM}
D. Achlioptas, P. Beame, and M. Molloy. {\em A sharp threshold in proof complexity yields lower bounds for satisfiability search.} {J. Comput. Syst. Sci.}, {\bf 68} (2), 238--268 (2004).

\bibitem{aco}
D. Achlioptas and A. Coja-Oghlan.
\newblock Algorithmic barriers from phase transitions.
\newblock In {\em 49th Annual IEEE Symposium on Foundations of Computer Science,
               FOCS 2008, October 25-28, 2008, Philadelphia, PA, USA}, pages 793--802. IEEE Computer Society, 2008.

\bibitem{aminfede}
D. Achlioptas, A. Coja-Oghlan and F. Ricci-Tersenghi
{\em On the solution-space geometry of random constraint satisfaction problems.}
{Random Struct. Algorithms},
{\bf 38} (3),
251-268 (2011).

\bibitem{fede}
D. Achlioptas and F. Ricci-Tersenghi
{\em Random Formulas Have Frozen Variables.}
SIAM J. Comput.,
{\bf 39} (1),
260-280 (2009).


\bibitem{berle} E.R. Berlekamp, R.J. McEliece, and H.C.A. van Tilborg. {\em On the inherent intractability of certain
coding problems.} IEEE Trans. Inform. Theory {\bf 24}, 384–386 (1978).

\bibitem{bb} B. Bollob\'{a}s. {\em A probabilistic proof of an asymptotic formula for the number of labelled regular graphs.} Europ. J. Combinatorics {\bf 1} 311-316 (1980).
  
\bibitem{cm}
H.S. Connamacher and M. Molloy.
\newblock The satisfiability threshold for a seemingly intractable random
  constraint satisfaction problem.
\newblock {\em SIAM J. Discrete Math.}, 26(2):768--800, 2012.

\bibitem{cc} C. Cooper. {\em The cores of random hypergraphs with a given degree sequence.}
 Random Structures Algorithms {\bf 25} (2004) 353~-~375.

\bibitem{daud}
H. Daud{\'e}, M. M{\'e}zard, T. Mora, and R. Zecchina.
{\em Pairs of SAT Assignments and Clustering in Random Boolean Formulae.}
Theoretical Computer Science,
{\bf 393} (1-3),
260-279 (2008).

\bibitem{cuc}  
M. Dietzfelbinger, A. Goerdt, M. Mitzenmacher, A. Montanari, R. Pagh and M. Rink.
Tight thresholds for cuckoo hashing via XORSAT. 
In {\em Automata, Languages and Programming, 37th International
               Colloquium, ICALP 2010, Bordeaux, France, July 6-10, 2010,
Part I}, pages 213-225. Springer, 2010.


\bibitem{dub}
O. Dubois and J. Mandler.
\newblock The 3-XORSAT threshold.
\newblock In {\em 43rd Symposium on Foundations of Computer Science (FOCS
               2002), 16-19 November 2002, Vancouver, BC, Canada}, pages 769-778. IEEE Computer Society, 2002.

\bibitem{young}
M. Guidetti and A.P. Young. {\em Complexity of several constraint-satisfaction problems using the heuristic classical algorithm WalkSAT.} Phys. Rev. E, {\bf 84} (1), 011102, July 2011.

\bibitem{ikkm}
M. Ibrahimi, Y. Kanoria, M. Kraning, and A. Montanari.
\newblock The set of solutions of random xorsat formulae.
\newblock In {\em Proceedings of the Twenty-Third Annual ACM-SIAM Symposium
               on Discrete Algorithms, SODA 2012, Kyoto, Japan, January
               17-19, 2012}, pages 760--779. SIAM, 2012.

\bibitem{jlr}
S. Janson, T. {\L}uczak and A. Ruci{\'n}ski.
{Random Graphs.} Wiley, New York (2000).

\bibitem{jhk} J.H.Kim.  {\em Poisson cloning model for random graphs.}  	arXiv:0805.4133v1

\bibitem{tlcomp}  T. {\L}uczak.  {\em Component behaviour near the critical point of the random graph process.}
Rand. Struc. \& Alg. {\bf 1} (1990), 287~-~310.


\bibitem{mmbook}
M. M{\'e}zard and A. Montanari.
\newblock {\em Information, Physics, and Computation}.
\newblock Oxford University Press, Inc., New York, NY, USA, 2009.

\bibitem{PhysRevLett.94.197205}
M. M{\'e}zard, T. Mora,  and R. Zecchina
{\em Clustering of Solutions in the Random Satisfiability Problem.}
Phys. Rev. Lett.,
{\bf 94} (19),
197205 (2005).

\bibitem{sp}
M. M{\'e}zard, G. Parisi, and R. Zecchina
{\em Analytic and Algorithmic Solution of Random Satisfiability Problems.}
Science,
{\bf 297},
812~-~815 (2002).


\bibitem{mez}
M. M{\'e}zard, F. Ricci-Tersenghi, and R. Zecchina. {\em Alternative solutions to diluted $p$-spin models and XORSAT
problems.} J. Stat. Phys. {\bf 111}, 505~-~533, (2003).

\bibitem{mmcore} M. Molloy {\em Cores in random hypergraphs and boolean formulas.}
Random Structures and Algorithms {\bf 27}, 124~-~135 (2005).

\bibitem{mr} M. Molloy and B. Reed. {\em A critical point for random graphs with a given degree sequence.}
Random Structures and Algorithms {\bf 6} 161~-~180 (1995).

\bibitem{ms} M. Molloy and M. Salavatipour. {\em The resolution complexity of random constraint satisfaction problems.} SIAM J. Comp. {\bf 37}, 895~-~922 (2007).

\bibitem{pittel} B. Pittel. {\em
On trees census and the giant component  in  sparse random graphs.}
Random Structures and Algorithms {\bf 1} (1990), 311~-~342.

\bibitem{psw} B. Pittel, J. Spencer and N. Wormald.
{\em Sudden emergence of a giant $k$-core in a random graph.}
J. Comb. Th. B {\bf 67}, 111~-~151 (1996).

\end{thebibliography}
\end{document}